\documentclass[11pt,a4paper,oneside]{article}
\usepackage[top=3cm, bottom=3cm, left=2cm, right=2cm]{geometry}
\usepackage{amsthm,amsmath,amsfonts,amssymb}
\usepackage[]{natbib}
\usepackage{graphicx}
\usepackage[T2A,T1]{fontenc}
\usepackage[utf8]{inputenc}
\usepackage[russian,english]{babel}
\usepackage[dvipsnames]{xcolor}
\usepackage{amsmath}
\usepackage{mathrsfs}
\usepackage{mathtools}
\usepackage{bbm}
\usepackage{verbatim}
\usepackage{algorithm}
\usepackage{dsfont}
\usepackage{bbm}
\RequirePackage[colorlinks,citecolor=blue,urlcolor=blue,backref=page,backref=page, linkcolor=blue]{hyperref}
\usepackage{multirow}
\usepackage[noend]{algpseudocode}
\usepackage{dsfont}
\usepackage{bm}
\usepackage{algorithmicx}
\usepackage{siunitx}
\usepackage{adjustbox}
\def\simind{\stackrel{\mbox{\footnotesize ind}}{\sim}}
\def\simiid{\stackrel{\mbox{\footnotesize iid}}{\sim}}

\def\Pr{\text{Pr}}
\usepackage{textcomp}
\usepackage{subfigure}

\usepackage{amsthm,amsmath,amssymb,mathrsfs,dsfont,mathtools}
\usepackage{bm}
\usepackage{graphicx}
\usepackage[colorinlistoftodos]{todonotes}
\usepackage{hyperref}
\usepackage{booktabs}
\usepackage[all]{nowidow}
\usepackage{setspace}
\usepackage{algorithm}
\usepackage[noend]{algpseudocode}
\usepackage{tikz} 
\usepackage[sf]{titlesec}
\usepackage{sectsty}
\allsectionsfont{\centering \normalfont\scshape} 
\usepackage{lipsum}
\usepackage{adjustbox}
\usepackage{lineno} 
\theoremstyle{plain}

\theoremstyle{definition}

\theoremstyle{remark}

\usepackage{authblk}

\usepackage{graphicx}%
\usepackage{multirow}%
\usepackage{amsmath,amssymb,amsfonts}%
\usepackage{amsthm}%
\usepackage{mathrsfs}%
\usepackage[title]{appendix}%
\usepackage{xcolor}%
\usepackage{textcomp}%
\usepackage{manyfoot}%
\usepackage{booktabs}%
\usepackage{algorithm}%
\usepackage{algorithmicx}%
\usepackage{algpseudocode}%
\usepackage{listings}%
\usepackage{bbm}
\usepackage{comment}
\usepackage{dirtytalk}

\usepackage{colortbl} 
\usepackage{adjustbox} 

\usepackage{bm}
\usepackage{dsfont}
\def\simind{\stackrel{\footnotesize{\mbox{ind}}}{\sim}}
\def\simiid{\stackrel{\footnotesize{\mbox{iid}}}{\sim}}
\def\btheta{\bm{\theta}}
\def\bSigma{\bm{\Sigma}}
\def\tr{\text{tr}}
\def\drm{\textrm{d}}

\def\model{\textsc{co}$^3$ }
\def\modelnosp{\textsc{co}$^3$}
\def\SMnsp{Supplementary Material}

\usepackage{dutchcal}

\def\p{\mathcal{p}}

\title{A Bayesian Model for Co-clustering Ordinal Data\\ 
with Informative Missing Entries}
\date{}
\author[1]{Alice Giampino}
\author[2]{Antonio Canale} 
\author[1]{Bernardo Nipoti}

\affil[1]{Department of Economics, Management and Statistics, University of Milano-Bicocca, Piazza dell'Ateneo Nuovo, 1, Milan, 20126, Italy.}
\affil[2]{Department of Statistics, University of Padova, Via C. Battisti, 241, Padova, 35121, Italy.}

\begin{document}

\maketitle

\abstract{Several approaches have been proposed in the literature for clustering multivariate ordinal data. These methods typically treat missing values as absent information, rather than recognizing them as valuable for profiling population characteristics. To address this gap, we introduce a Bayesian nonparametric model for co-clustering multivariate ordinal data that treats censored observations as informative, rather than merely missing. We demonstrate that this offers a significant improvement in understanding the underlying structure of the data. Our model exploits the flexibility of two independent Dirichlet processes, allowing us to infer potentially distinct subpopulations that characterize the latent structure of both subjects and variables. The ordinal nature of the data is addressed by introducing latent variables, while a matrix factorization specification is adopted to handle the high dimensionality of the data in a parsimonious way. The conjugate structure of the model enables an explicit derivation of the full conditional distributions of all the random variables in the model, which facilitates seamless posterior inference using a Gibbs sampling algorithm. We demonstrate the method's performance through simulations and by analyzing politician and movie ratings data.}

\maketitle

\section{Introduction}\label{sec1}

Multivariate ordinal data, consisting of repeated measurements of vectors of ordinal responses, play a crucial role in various fields. Our focus is on scenarios where repeated measures are available for only a subset of the vector entries, and the missing data can be viewed as informative rather than purely random. For example, in the analysis of political votes, the voting records of $n$ politicians on $p$ legislative bills are only available for the sessions they attended, where absence from a session may represent a deliberate political strategy. Similarly, in recommendation systems, the data consists of customer ratings on $p$ items. In the well-known Netflix Prize data \citep{bennett2007netflix}, missingness arises when users choose not to rate certain movies, potentially reflecting their lack of interest, preference, or other underlying factors.

Our analysis of this type of data focuses on clustering both the statistical units and the entries of the ordinal random vectors. This problem can be framed within the context of co-clustering methods \citep{hartigan1972direct}, a set of techniques also known as biclustering or two-mode clustering. The core idea is to simultaneously cluster the rows and columns of an $(n\times p)$-dimensional data matrix $\bm{Y}$, with rows associated to $n$ statistical units and columns consisting of $p$ ordinal outcomes.
Over the past three decades, co-clustering methods have been widely applied across various fields \citep{busygin2008biclustering}, with some approaches resorting to Bayesian nonparametric tools \citep{meeds2007nonparametric,wang2011nonparametric,wang2012feature}. Despite extensive literature, most existing methods fail to adequately address the challenges posed by informative missingness. A notable exception is the \texttt{R} package \texttt{biclustermd} by \citet{reisner2019biclustermd}, which uses a geometric approach to optimally rearrange the rows and columns of the data matrix when missing values are present. To our knowledge, no model-based approach for co-clustering with missing entries has yet been explored in the literature. We fill this gap by introducing a flexible Bayesian model capable of performing co-clustering while accounting for the informative nature of missing data. This approach, referred to as the Co-Clustering model for Ordinal Censored Observations (\modelnosp), employs a latent variable representation to jointly model both the ordinal responses and the indicator variables encoding the presence of missing entries. 
\model builds on the idea of linking observed discrete data with latent continuous variables, thus following a well-known strategy in Bayesian modeling \citep{albert1993bayesian, Kot05, canale2011bayesian, kowal2020simultaneous}. 
At the level of latent variables, \model employs a Bayesian matrix factorization representation \citep{salakhutdinov2008bayesian}, exploiting the idea that the observed multivariate responses of a statistical unit are driven by a smaller set of unobserved latent factors. Additionally, the definition of \model relies on the flexibility of two independent Dirichlet processes, which allow the model to infer potentially distinct subpopulations characterizing the latent structure of both subjects and variables. This construction not only facilitates the clustering of statistical units and variable entries but also provides a robust framework for effectively addressing the complexities associated with informative censoring. Our investigations show that, while the model is specifically designed to account for informative missingness, its applicability extends to situations where data are missing at random as well.

The rest of the paper is organized as follows. Section \ref{sec:modelch2} is dedicated to the specification of the model. A strategy for posterior computation is presented in  Section \ref{sec:postch2}. The performance of the model is investigated by means of the analysis of synthetic and real data, as presented in Sections \ref{sec:simch2} and \ref{sec:datareal}, respectively. Additional results on the full conditional distributions for posterior sampling, along with further details on the simulation study and the real dataset analyses, are provided in the \SMnsp.

\section{A co-clustering model for
ordinal censored observations }\label{sec:modelch2}

We let $\bm{y}_i=(y_{i1},\ldots,y_{ip})$, with $i = 1, \dots, n$, be the $i$-th row of a data matrix $\bm{Y}$, that is a vector of ordinal responses with the general entry $y_{ij} \in [c]$, for $j =1, \dots, p$, where $[c]=\{1, \dots, c\}$. For example, in our two motivating applications, $y_{ij}$  represents the vote of the $i$-th senator in the $j$-th voting session, or the rating given by $i$-th user to the $j$-th movie. In both these applications, as well as in many other contexts, it is common for some components of each observation $\bm{y}_i$ to be missing, with the missingness occurring in a non-random manner.

 \subsection{Model formulation}
 \label{sec:modelformulation}
 
 We formalize the missingness by endowing each $\bm{y}_i$ with a vector $\bm{\delta}_i=(\delta_{i1},\ldots,\delta_{ip})$, with $\delta_{ij}$ indicating whether the $j$-th component $y_{ij}$ of $\bm{y}_i$ was observed ($\delta_{ij}=1$) or not ($\delta_{ij}=0)$. 
 This  can be easily generalized to the case where, for each unit, there exist $q$  ordinal responses, each accompanied by a corresponding missingness indicator. We will henceforth focus on the specific case with $q=1$.

Similarly to \citet{albert1993bayesian} or \citet{Kot05}, we link both the observed ordinal $y_{ij}$ and the missingness indicators  $\delta_{ij}$ with latent continous variables. Specifically, for the ordinal responses, we introduce $\bm{z}_i=(z_{i1},\ldots,z_{ip})$  such that
\begin{align*}
    y_{ij}&=\kappa&\text{ if } \gamma_{\kappa-1}<z_{ij}\leq \gamma_{\kappa};\\      
    y_{ij}&\in[c]&\text{ if } -\infty=\gamma_{0}<z_{ij}\leq \gamma_{c}=\infty.
\end{align*}
where $-\infty=\gamma_{0}<\gamma_{1}<\ldots<\gamma_{c-1}<\gamma_{c}=\infty$ are arbitrarily fixed cutoffs. 
Similarly, for the censoring variables, we introduce $\bm{w}_i=(w_{i1},\ldots,w_{ip})$  such that $\delta_{ij} = 1$ if and only if $w_{ij} \geq 0.$
This formulation leads to the following joint distribution for the observed data: 
\begin{equation}\label{eq:linkelihood}    
    \prod_{i,j : \delta_{ij} =1} \Pr\left(z_{ij} \in (\gamma_{y_{ij}-1}, \gamma_{y_{ij}}]\right).
\end{equation}

We now define a Bayesian factor  model for the joint distribution of the latent variables
$\bm{x}_{ij} = (z_{ij}, w_{ij})$.
Specifically, we introduce $\bm{\theta}_{1i}$ and $\bm{\theta}_{2j}$ to denote the $(d\times 2)$-dimensional factor matrices for the $i$-th individual and $j$-th response, respectively, with $d\ll n,p$. 
We further model the factor matrices $\bm{\theta}_{1i}$ and $\bm{\theta}_{2j}$ with independent Dirichlet processes (DP), leading to
\begin{align}\label{eq:modelgeneral}
    \bm{x}_{ij}\mid\btheta_1, \btheta_2 &\simind \text{N}_{2}(\btheta_{1i}^\intercal\btheta_{2j}, \bSigma),& i=1,\ldots,n;\, j=1,\ldots, p;\notag\\
    (\btheta_{1i},\btheta_{2j})\mid F_1,F_2 &\simiid F_1\times F_2,& i=1,\ldots,n;\, j=1,\ldots, p;\\[1pt]
    F_{l}&\simind \text{DP}(\alpha_{l},H_{l}),&l=1,2;\notag
\end{align}
where $\btheta_1=\{\btheta_{11},\ldots,\btheta_{1n}\}$, $\btheta_2=\{\btheta_{21},\ldots,\btheta_{2p}\}$,  $\bSigma$ is a variance-covariance matrix, and $\alpha_l$ and $H_l$ denote, respectively, the concentration parameter and base probability measure of the 
DP $F_l$. 
The base measures $H_l$ are specified to be matrix normal distributions \citep[see, e.g.,][for their use in the context of mixture models]{viroli2011finite} with mean matrix $\bm{M}_l$ of dimension $d\times 2$ and covariance matrices $\bm{U}_l$ and $\bm{V}_l$ of dimensions $2\times 2$ and $d\times d$, respectively. We henceforth assume that $\bSigma$ is diagonal with $(\sigma_1^2,\sigma_2^2)$ on the diagonal.

Notably, the introduction of a latent layer of continuous random variables considerably simplifies the task of writing the joint distribution of all the elements in the model, which is the starting point of next section.

\subsection{Model properties}\label{sec:postdisrtich2}

In view of the definition of a MCMC algorithm for posterior inference, described in Section \ref{sec:postch2}, we study the joint conditional distribution of the random elements that constitute the model in \eqref{eq:linkelihood} and \eqref{eq:modelgeneral}, given the data. More specifically, we show the derivation of the conditional distribution that is obtained after marginalizing with respect to the random probability measures $F_1$ and $F_2$. This step conveniently simplifies the task of posterior sampling by analytically integrating out the infinite-dimensional parameters of the model.

We observe that, given the almost sure discreteness of $F_1$, the random matrices $\btheta_1$ will display ties with positive probability and thus the set $\btheta_1$ can be equivalently described in terms of $k_n\leq n$ distinct values $\btheta_{1\ell_1}^*$, with $\ell_1=1,\ldots,k_n$, and their frequency $n_{\ell_1}=\sum_{i=1}^n\delta_{\btheta_{1\ell_1}^*}(\btheta_{1i})$ in $\btheta_1$. Similarly, $\btheta_2$ can be described by $k_p\leq p$ distinct values $\btheta_{2\ell_2}^*$, with $\ell_2=1,\ldots,k_p$, and their frequency $n_{\ell_2}=\sum_{j=1}^p\delta_{\btheta_{2\ell_2}^*}(\btheta_{2j})$ in $\btheta_2$. As a result, using independent Dirichlet processes to model $\btheta_1$ and $\btheta_2$ conveniently facilitates the simultaneous clustering of subjects and responses. This is particularly relevant in our motivating political application, where analysts are interested in grouping politicians based on their actual voting behaviors and in determining whether this alignment corresponds with their party affiliations. At the same time, this approach can help in categorizing voting sessions to identify patterns in legislative priorities and uncover strategic collaborations across party lines.\\
The distributions of $(n_{1},\ldots,n_{k_n})$ and $(p_{1},\ldots,p_{k_p})$ are characterized by the exchangeable partition probability function (EPPF) of the DP, for which we have 
\begin{equation*}
            \Pi_{k_n}^{(n)}(n_1,\ldots,n_{k_n})=\frac{\alpha_1^{k_n}}{(\alpha_1)_n}\prod_{\ell_1=1}^{k_n} (n_{\ell_1}-1)!, \quad 
            \Pi_{k_p}^{(p)}(p_1,\ldots,p_{k_p})=\frac{\alpha_2^{k_p}}{(\alpha_2)_p}\prod_{\ell_2=1}^{k_p} (p_{\ell_2}-1)!,
\end{equation*}
where the symbol $(a)_n=a(a+1)\cdots(a+n-1)$ is used to denote the ascending factorial. To be more specific, $\Pi_{k_n}^{(n)}(n_1,\ldots,n_{k_n})$ gives the probability of observing any specific partition of the elements of $\btheta_1$ in $k_n$ distinct values of cardinality $\{n_1,\ldots,n_{k_n}\}$; similarly for $\Pi_{k_p}^{(p)}(p_1,\ldots,p_{k_p})$. 

We introduce $\btheta=\{\btheta_1,\btheta_2\}$ and $\bm{X}$ to be the tensor with $n$ rows, $p$ columns, and $2$ tubes, containing in row $i$, column $j$, the vector $\bm{x}_{ij}$ and we focus on  studying the conditional distribution $\p(\btheta,\bm{X}\mid \mathcal{D})$ of the latent variables $\{\btheta,\bm{X}\}$, given the data $\mathcal{D}$. Here and henceforth, $\p(\cdot)$ denotes the distribution of a generic random element.
Marginalizing the distribution implied by \eqref{eq:linkelihood} and \eqref{eq:modelgeneral} with respect to the DPs $F_1$ and $F_2$, we get
\begin{align}\label{eq:joint}
    \p(\btheta,\bm{X}\mid \mathcal{D}) &\propto \Pi_{k_n}^{(n)}(n_1,\ldots,n_{k_n}) \Pi_{k_p}^{(p)}(p_1,\ldots,p_{k_p})\prod_{\ell_1=1}^{k_n}h_{1}(\btheta_{1\ell_1}^*)\prod_{\ell_2=1}^{k_p} h_{2}(\btheta_{2\ell_2}^*)\notag\\
        &\times \prod_{i\in\mathcal{C}_{1\ell_1}}\prod_{j\in\mathcal{C}_{2\ell_2}} 
        \p(y_{ij},\delta_{ij}\mid \bm{x}_{ij})\p(\bm{x}_{ij}\mid \btheta_{1\ell_1}^*,\btheta_{2\ell_2}^*),
\end{align}
where $h_l$ denotes the probability density function corresponding to the base measure $H_l$, for $l=1,2$, and $\mathcal{C}_{1\ell_1}=\{i\in\{1,\ldots,n\}\,:\,\bm{\theta}_{1i}=\bm{\theta}_{1\ell_1}^*\}$ and $\mathcal{C}_{2\ell_2}=\{j\in\{1,\ldots,p\}\,:\,\btheta_{2j}=\btheta_{2\ell_2}^*\}$. From \eqref{eq:joint} one can obtain the full conditional distributions of the elements of $\btheta$ and $\bm{X}$. 

Given the focus is on simultaneously clustering of subjects and responses, it is interesting to study the prior properties of $k=k_n k_p$, which we refer to as the number of bivariate clusters and which corresponds to the cardinality of the set of all possible pairs of cluster assignments $\{(\ell_1,\ell_2)\,:\,\ell_1=1,\ldots,k_n \text{ and }\ell_2=1,\ldots,k_p\}$. The indepencence of $F_1$ and $F_2$ facilitates the task, and we get that $k$ has distribution 
\begin{equation*}
   \Pr(k = \ell )=
    \frac{1}{(\alpha_1)_n(\alpha_2)_p}\sum_{i,j: ij = \ell} \alpha_1^i \alpha_2^j |s(n, i)||s(p, j)|,
\end{equation*}
where $s(n, i)$ is the Stirling number of the first type. 
\begin{figure}[t]
    \centering
    \includegraphics[scale=0.3]{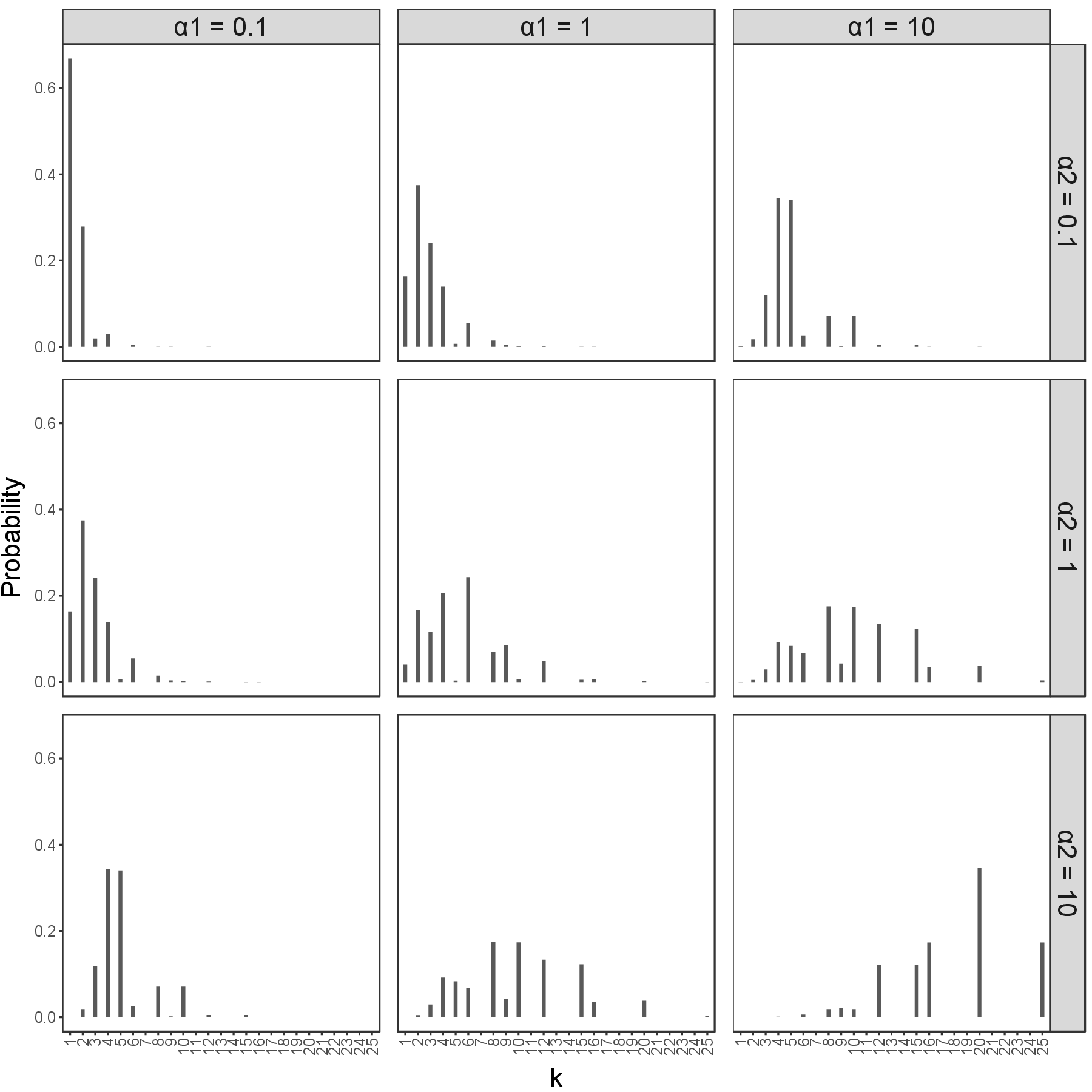}
    \caption{Prior distribution of the number of bivariate clusters $k$ in \modelnosp, for $n=p=5$, and different values of $\alpha_1$ and $\alpha_2$ ranging in $\{0.1,1,10\}$.}
    \label{fig:k_mass_vert}
\end{figure}
As an example, Figure \ref{fig:k_mass_vert} displays the prior distribution of the number of bivariate clusters $k$ when $n=p=5$ and for different values of the concentration parameters $\alpha_1$ and $\alpha_2$. As the latter ones  increase, the distribution of $k$ concentrates on larger values. Consistently with this, the expected value of $k$ is
\begin{equation}\label{eq:exp_k}
    \mathbbm{E}[k]=\alpha_1\alpha_2\sum_{i=1}^n\frac{1}{\alpha_1+i-1}\sum_{j=1}^p\frac{1}{\alpha_2+j-1}. 
\end{equation}
Finally, we observe that, as both $n$ and $p$ grow to infinity, the number of bivariate clusters grows proportionally to $\log(n)\log(p)$. Specifically, we have
$$\frac{k}{\log(n)\log(p)}\longrightarrow \alpha_1\alpha_2\; \text{ almost surely},$$
as $n,p\longrightarrow\infty$.

\section{Posterior Computation}\label{sec:postch2}

Equation \eqref{eq:joint} provides the starting point to devise a Gibbs sampler for posterior inference. The update of the parameters is conveniently facilitated by the availability of closed-form full conditional distributions, which we discuss for the random elements of model \eqref{eq:linkelihood} and \eqref{eq:modelgeneral}, namely $\bm{z}_i,\bm{w}_i,\btheta_{1i},\btheta_{2i},\sigma_1^2,\sigma_2^2$. To this end, we introduce additional notation. For any $r=1,2$, we denote the $r$-th column of 
$\btheta_{1i}$ and $\btheta_{2j}$ as $\btheta_{1i}^{(r)}$ and $\btheta_{2j}^{(r)}$, respectively. Moreover, we introduce the $(d\times n)$-dimensional matrix $\btheta_1^{(r)}$ as the matrix whose $i$-th column coincides with $\btheta_{1i}^{(r)}$, and the $(d\times p)$-dimensional matrix $\btheta_2^{(r)}$ as the matrix whose $j$-th column coincides with $\btheta_{2j}^{(r)}$. 

The full conditional distributions of $\bm{z}_i$ and $\bm{w}_{i}$ 
are $p$-dimensional truncated multivariate normal distributions with independent components. This implies that, for any $j=1,\ldots,p$,
\begin{equation}\label{eq:fc_zij}
    z_{ij}\mid \ldots \simind \delta_{ij}\textsc{TN}\left(\btheta_{2j}^{(1)\intercal}\btheta_{1i}^{(1)},\sigma_1^2;\gamma_{y_{ij}-1},\gamma_{y_{ij}}\right)+(1-\delta_{ij})\textsc{N}\left(\btheta_{2j}^{(1)\intercal}\btheta_{1i}^{(1)},\sigma_1^2\right),
\end{equation}
where $\textsc{TN}(m,s^2;a,b)$ is used to denote a two-sided truncated normal with mean $m$, variance $s^2$ and support $(a,b)$. Similarly, 
\begin{equation}\label{eq:fc_wij}
    w_{ij}\mid\ldots\simind \delta_{ij}\textsc{TN}\left(\btheta_{2j}^{(2)\intercal}\btheta_{1i}^{(2)},\sigma_2^2;0,\infty\right)+(1-\delta_{ij})\textsc{TN}\left(\btheta_{2j}^{(2)\intercal}\btheta_{1i}^{(2)},\sigma_2^2;-\infty,0\right).
\end{equation}
Next, we observe that, conditionally on the continuous latent vectors $\bm{z}_i$ and $\bm{w}_i$, the random vectors $\btheta_{1i}$ and $\btheta_{2i}$ are independent of the observations $\bm{y}_i$ and $\bm{\delta}_i$. Their distribution is governed by model \eqref{eq:modelgeneral}, which incorporates two independent DPs. The literature on computational methods for DP-based models, or other discrete nonparametric priors, is extensive, providing various strategies to handle the infinite dimensionality of such models. Examples include \citet{escobar1995bayesian}, \citet{Pap08}, \citet{Kal11}, and \citet{ICS}. In this work, we adapt the marginal approach of \citet{escobar1995bayesian} to accommodate the additional complexity introduced by the two DPs in our model. Specifically, the marginalization of model \eqref{eq:modelgeneral} with respect to $F_1$ and $F_2$, yields full conditional distributions for $\btheta_{1i} = (\btheta_{1i}^{(1)}, \btheta_{1i}^{(2)})$ and $\btheta_{2j} = (\btheta_{2j}^{(1)}, \btheta_{2j}^{(2)})$ whose form is reminiscent of the generalized P\'olya urn scheme \citep{Bla73}. Namely,
\begin{align}\label{eq:fc_theta_1}
   \Pr(\btheta_{1i}\in \drm \bm{t}\mid \ldots)&=\pi_{1i0}G_{1i}(\drm \bm{t})+\sum_{\ell=1}^{k_{n(i)}}\pi_{1i\ell}\delta_{\btheta_{1\ell(i)}^*}(\bm{t})& i=1,\ldots,n\\
\label{eq:fc_theta_2}
   \Pr(\btheta_{2j}\in \drm \bm{t}\mid \ldots)&=\pi_{2j0}G_{2j}(\drm \bm{t})+\sum_{\ell=1}^{k_{p(j)}}\pi_{2j\ell}\delta_{\btheta_{2\ell(j)}^*}(\bm{t})& j=1,\ldots,p
\end{align}
where the subscript $(i)$ (or $(j)$) is used to denote quantities that are computed after excluding the element $\btheta_{1i}$ from $\btheta_{1}$ (or $\btheta_{2j}$ from $\btheta_2$). The weights $\{\pi_{1i0},\ldots,\pi_{1ik_{n(i)}}\}$ in \eqref{eq:fc_theta_1} and $\{\pi_{2j0},\ldots,\pi_{2jk_{p(j)}}\}$ in \eqref{eq:fc_theta_2}, provided in 
the \SMnsp, are available in closed form. The distributions $G_{1i}$ in \eqref{eq:fc_theta_1} and $G_{2j}$ in \eqref{eq:fc_theta_2} are easy to sample from as they can be written as a conditional chain of $d$-dimensional normal distributions, as described in 
the \SMnsp. A convenient simplification that will be implemented for the analyses of Sections \ref{sec:simch2} and \ref{sec:datareal} is achieved by assuming that the base measures $H_l$ are characterized by independence between columns. This is equivalent to assuming that the column covariance matrices $\bm{U}_l$ are diagonal, which allows us to decompose the matrix variate normal base measure into a product of independent multivariate normal distributions. As a result, the form of the weights $\pi_{1i0}$ and $\pi_{2j0}$ simplifies considerably, and the problem of sampling from $G_{1i}$ and $G_{2j}$ reduces to sampling from independent $d$-dimensional normal distributions. Again, details are deferred to the \SMnsp.\\ 
It is well known that algorithms based on P\'olya urn schemes can suffer of slow mixing \citep[see, e.g., the discussion in][]{ishwaran2001gibbs}. A solution to deal with this problem is the introduction of a reshuffling step to update the distinct values of the latent variables
from their full conditional distributions. We observe that 
\begin{align}\label{eq:acc}
    \Pr( \btheta_{1\ell_1}^\ast \in (\drm\bm{t}_1, \drm\bm{t}_2)\mid \ldots)
    &\propto H_1(\drm\bm{t}_1, \drm\bm{t}_2)\notag\\
    &\times \exp\Big\{-\frac{1}{2} \tr\big[\frac{1}{\sigma_1^2}\sum_{i \in \mathcal{C}_{\ell_1}} (\bm{z}_{i}-\btheta_2^{(1)\intercal}\bm{t}_1)(\bm{z}_{i}-\btheta_2^{(1)\intercal}\bm{t}_1)^\intercal \big]\Big\}\notag\\
    &\times \exp \Big\{-\frac{1}{2} \tr\big[\frac{1}{\sigma_2^2}(\bm{w}_{i}-\btheta_2^{(2)\intercal} \bm{t}_2)(\bm{w}_{i}-\btheta_2^{(2)\intercal} \bm{t}_2)^\intercal \big] \Big\},
\end{align}
for any $\ell_1=1,\ldots,k_n$, and that, similarly,
\begin{align}\label{eq:acc2}
    \Pr( \btheta_{2\ell_2}^\ast \in (\drm\bm{s}_1, \drm\bm{s}_2)\mid \ldots) 
    &\propto H_2(\drm\bm{s}_1, \drm\bm{s}_2)\notag\\
    &\times \exp\Big\{-\frac{1}{2} \tr\big[\frac{1}{\sigma_1^2}\sum_{j \in \mathcal{C}_{\ell_2}} (\bm{z}_{j}-\btheta_1^{(1)\intercal}\bm{s}_1)(\bm{z}_{j}-\btheta_1^{(1)\intercal}\bm{s}_1)^\intercal \big]\Big\}\notag\\
    &\times \exp \Big\{-\frac{1}{2} \tr\big[\frac{1}{\sigma_2^2}(\bm{w}_{j}-\btheta_1^{(2)\intercal} \bm{s}_2)(\bm{w}_{j}-\btheta_1^{(2)\intercal} \bm{s}_2)^\intercal \big] \Big\},
\end{align}
for any $\ell_2=1,\ldots,k_p$. As already observed for the distributions $G_{1i}$ and $G_{2j}$, also the distributions in \eqref{eq:acc} and \eqref{eq:acc2} can be written as a conditional chain of $d$-dimensional normal distributions. 
Moreover, under the additional assumption that the column covariance matrices $\bm{U}_l$ are diagonal, one gets that the columns of $\btheta_{1\ell_1}^*$ distributed as \eqref{eq:acc}, and similarly those of $\btheta_{1\ell_1}^*$ distributed as in \eqref{eq:acc2}, are independent $d$-dimensional normal random vectors, with parameters reported in the \SMnsp.\\ 
Finally, if the model is completed by specifying Inverse-gamma hyperpriors for the  $\sigma_1^2$ and $\sigma_2^2$, namely    $\sigma_l^2 \sim \text{IG}(\alpha_{\sigma_l}, \beta_{\sigma_l})$ with $l=1,2$,
then, the full conditionals for $\sigma_1^2$ and $\sigma_2^2$ are given by
\begin{align}\label{eq:fc_sigma1}
    \sigma_1^2\mid \ldots &\sim \text{IG}\left(\alpha_{\sigma_1}+\frac{np}{2}, \beta_{\sigma_1}+\frac{1}{2}\sum_{i=1}^n\sum_{j=1}^p (z_{ij} -\btheta_{1i}^{(1)\intercal  }\btheta_{2j}^{(1)})^2\right),\\
    \label{eq:fc_sigma2}\sigma_2^2\mid \ldots &\sim \text{IG}\left(\alpha_{\sigma_2}+\frac{np}{2}, \beta_{\sigma_2}+\frac{1}{2}\sum_{i=1}^n\sum_{j=1}^p (w_{ij} -\btheta_{1i}^{(2) \intercal} \btheta_{2j}^{(2)})^2\right).
\end{align}

Algorithm \ref{gibbs} outlines the steps of the Gibbs sampler we obtain by combining the sequential updates of the model parameters according to the corresponding full conditional distributions, with the reshuffling step. Extending the algorithm to the case of $2q$ response variables poses no significant challenges, as the additional latent factor vectors corresponding to the continuous latent variables are jointly handled within the two P\'olya urn steps of the algorithm. Finally, we note that throughout this work, the rows and the columns of the data matrix $\bm{Y}$ are partitioned, based on the posterior output produced by Algorithm \ref{gibbs}, by resorting to Variation of Information with complete linkage \citep{wade2018bayesian}.

\begin{algorithm}[h!]
\caption{Gibbs sampling for \modelnosp}\label{gibbs}
{\small
\begin{algorithmic}[1]
\State $\textbf{set} \text{ admissible initial values for the latent vectors } \btheta_1 \text{ and } \btheta_2$
\For {$\text{each iteration } b = 1\ldots,B$}:
\For {$\text{each } i=1, \ldots, n \text{ and } j=1, \ldots, p$}: \Comment{generalized P\'olya urns}
\State $\textbf{sample } \btheta_{1i} \text{ from Equation \ref{eq:fc_theta_1}}$
\State $\textbf{sample } \btheta_{2j} \text{ from {Equation \ref{eq:fc_theta_2}}}$
\EndFor
\State $\textbf{set } \btheta_{1}^{\ast} = \left(\btheta_{11}, \ldots, \btheta_{1 k_n}\right)$ be the vector of distinct parameters in  $\btheta_{1}$.
\State $\textbf{set } \btheta_{2}^{\ast} = \left(\btheta_{21}, \ldots, \btheta_{1 k_p}\right)$ be the vector of distinct parameters in  $\btheta_{2}$.
\For {$\text{each } \ell_1=1, \ldots, k_n$}: \Comment{reshuffling step 1}
\State $\textbf{let } C_{\ell_1} \text{ be the set of indexes } i \text{ such that } \btheta_{1i} = \btheta_{1\ell_1}^{\ast};$
\State $\textbf{sample } \btheta_{1\ell_1}^{\ast} \text{ from {Equation \ref{eq:acc}}}$
\EndFor
\For {$\text{each } \ell_2=1, \ldots, k_p$}: \Comment{reshuffling step 2}
\State $\textbf{let } C_{\ell_2} \text{ be the set of indexes } j \text{ such that } \btheta_{2j} = \btheta_{2\ell_2}^{\ast};$
\State $\textbf{sample } \btheta_{2\ell_2}^{\ast} \text{ from {Equation \ref{eq:acc2}}}$
\EndFor
\For {$\text{each } i=1, \ldots, n \text{ and } j=1, \ldots, p$}: \Comment{update of continuous latent variables}
\State $\textbf{sample } z_{ij} \text{ from Equation \ref{eq:fc_zij}}$ 
\State $\textbf{sample } w_{ij} \text{ from Equation \ref{eq:fc_wij}}$
\EndFor
\State $\textbf{sample } \sigma_1^2 \text{ from Equation \ref{eq:fc_sigma1}}$\Comment{update of hyperparameters}
\State $\textbf{sample } \sigma_2^2 \text{ from Equation \ref{eq:fc_sigma2}}$
\EndFor
\State $\textbf{end}$
\end{algorithmic}}
\end{algorithm}

A key step in implementing the \model model involves specifying the latent dimension $d$. A suitable choice of $d$ should strike a balance: on the one hand, a smaller $d$ is appealing as it keeps computations tractable and favours the interpretability of the analysis results. On the other hand, $d$ needs to be large enough to capture and distinguish the key features of the $p$-dimensional observations, even when projected onto a $d$-dimensional space, with $d \ll p$. We propose selecting $d$ on a case-by-case basis by comparing the predictive performance of models with different values of $d$. This will be done by evaluating the Log Pseudo Marginal Likelihood (LPML) \citep{geisser1979predictive} for a range of values of $d$. Our approach offers two main advantages: first, we keep the latent variables dimension fixed when implementing Algorithm \ref{gibbs}, thus avoiding the issue of dealing with trans-dimensional steps; second, it provides useful insights into how the choice of $d$ affects the model's predictive ability. 
Other strategies are possible and have been explored in recent literature. A notable approach, which has gained considerable attention in factor models, involves the use of shrinkage priors \citep{Bha11,Leg20}. These priors for $d$ have support on the set of positive integers and promote sparsity by inducing posterior shrinkage towards smaller dimensions.

\section{Simulation studies}\label{sec:simch2}

We assess the performance of \model in co-clustering individuals and items using synthetic data generated across various scenarios, differing in dimensionality and characterised by censoring mechanisms of varying intensity and nature. In each scenario, data are simulated under the correct model specification, specifically by generating the latent variables $\mathbf{x}_{ij}$ according to the first equation in \eqref{eq:modelgeneral}. The study is organized in two parts. In both, we assess the co-clustering performance of \model and we compare it against that one of the \texttt{biclustermd} R package. \model is implemented by running Algorithm \ref{gibbs} for 5,000 iterations, with the first half discarded as burn-in. For simplicity, rather than assigning hyperpriors as in \eqref{eq:fc_sigma1} and \eqref{eq:fc_sigma2}, we set $\sigma_1^2=0.1$ and $\sigma_2^2=1.5$. The concentration parameters $\alpha_1$ and $\alpha_2$ of $F_1$ and $F_2$ are both fixed equal to 1. Additionally, for $l=1,2$, we set $u_{l11}=u_{l22}=1/\sqrt{d}$, where $u_{l11}$ and $u_{l22}$ are the entries of the diagonal matrix $\bm{U}_l$, and $\bm{M}_l$ is defined as a matrix of zeros. While \model infers the number of clusters for both rows and columns of $\bm{Y}$, for \texttt{biclustermd} we set the number of clusters to match the actual numbers used in data simulation, thus facilitating row and column clustering for this alternative approach.

In the first part of this study, ordinal observations are generated to resemble the movie ratings data analysed in Section \ref{movie}. Specifically, we assume that the ratings $\bm{y}_i$ take values in $\{1,2,3\}$ and consider a scenario characterised by three types of users and three types of movies. To achieve this, the matrices $\btheta_{1i}$ and $\btheta_{2j}$ are randomly generated from mixtures with three components each. Additionally, we censor 5\% of the observations, randomly selected from records corresponding to the lowest ratings, thereby simulating a mechanism where missing entries may indicate a lack of interest in specific movies. We henceforth refer to this mechanism as informative censoring. Data are generated with different values for $n$ and $p$, with $(n,p) \in \{(50,50), (100,100), (200,200)\}$. For each scenario, 100 independent datasets are generated. An example is provided in Figure \ref{fig:ordinal} in the \SMnsp.

\begin{figure}
    \centering
    \includegraphics[clip,trim=1cm 0.5cm 3.5cm 0.5cm,scale=0.2]{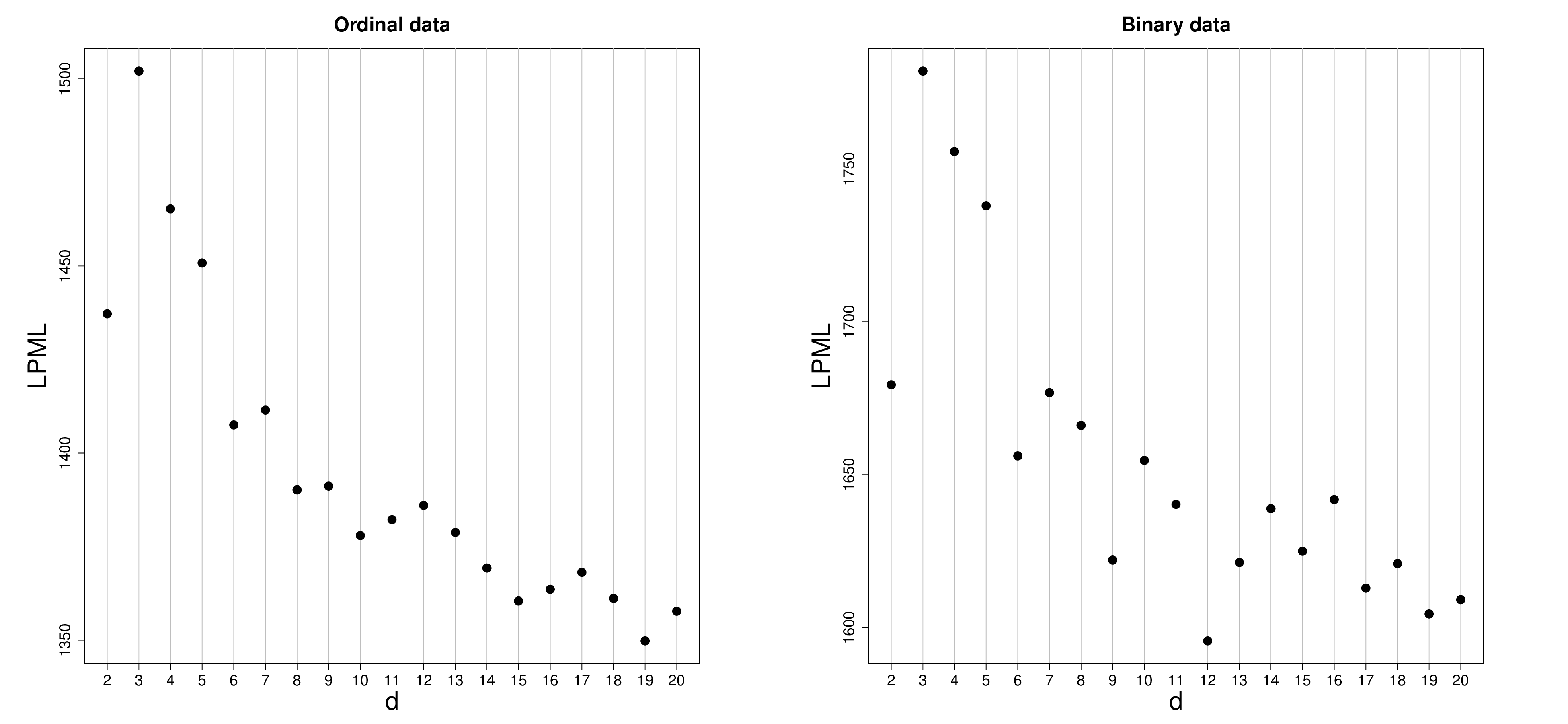}
     \caption{Simulated data with informative censoring. LPML for different values of the latent dimension $d$ on a randomly selected dataset with $n=p=50$ and $5\%$ of informative censoring, with ordinal observations (left panel) and binary observations (right panel).}\label{fig:lpml_joint}
\end{figure}

In order to select $d$, we ran the model on a single dataset, selected at random from the 100 replicates with $n = p = 50$ and 5\% censored observations, for different specifications of $d \in \{2, \ldots, 20\}$. The LPML plot, shown in the left panel of Figure \ref{fig:lpml_joint}, suggests that $d = 3$ is optimal, and this value is fixed for all subsequent analyses.

Given the bivariate nature of the co-clustering problem, the performance of a method is assessed using a bivariate extension of the Adjusted Rand Index \citep[ARI,][]{rand1971objective}, referred to as the bivariate ARI (BARI), to compare true and estimated partitions. The BARI provides a summary measure of a model's ability to correctly cluster both rows and columns of a data matrix. Specifically, when analysing the data matrix $\bm{Y}$, the BARI is defined as the ARI between the estimated and true partition of the $np$ data entries $\{y_{ij} \,:\, i=1,\ldots,n \text{ and }j=1,\ldots,p\}$, where $y_{i_1j_1}$ and $y_{i_2j_2}$, with $(i_1,j_1)\neq(i_2,j_2)$, are in the same cluster if and only if the $i_1$-th and $i_2$-th rows belong to the same cluster, and the $j_1$-th and $j_2$-th columns belong to the same cluster, according to the marginal partitions of rows and columns.
\begin{figure}
    \centering
    \includegraphics[clip,trim=0cm 3.2cm 0cm 3.2cm,scale=0.4]{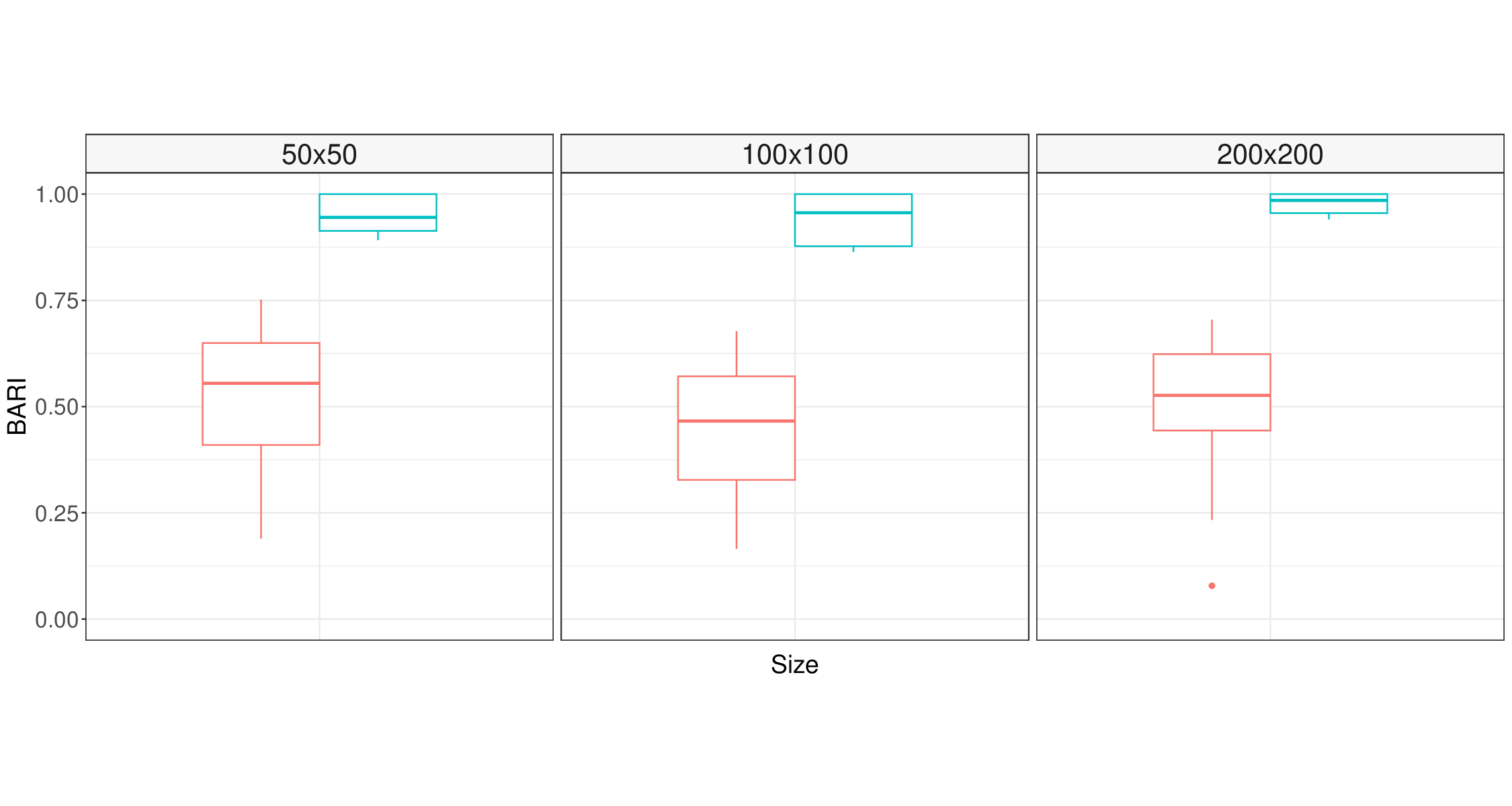}
    \caption{Simulated ordinal data with informative censoring. Boxplot for the BARI comparing the true bivariate partition and those identified by \texttt{biclustermd} (left boxplots) and \model (right boxplots), for different data size.}\label{fig:BRI_sc1_joint}
\end{figure}
The results of our study are presented in Figure \ref{fig:BRI_sc1_joint}, showing that \model consistently outperforms \texttt{biclustermd}. The performance of \model appears rather stable performance across varying dataset sizes, with the median BARI increasing slightly as dataset size grows, indicating that larger datasets favour the recovery of true latent clusters. In contrast, \texttt{biclustermd} yields BARI values consistently around 50\%. We also evaluated \modelnosp's ability to estimate the marginal partitions for the rows and columns of $\bm{Y}$. The results of our analysis, summarized in Table \ref{tab:mari_sc1} in the \SMnsp, confirm \modelnosp's robust performance across different dataset sizes.

\begin{figure}
    \centering
    \includegraphics[clip,trim=0cm 0cm 0cm 0cm,height=6cm]{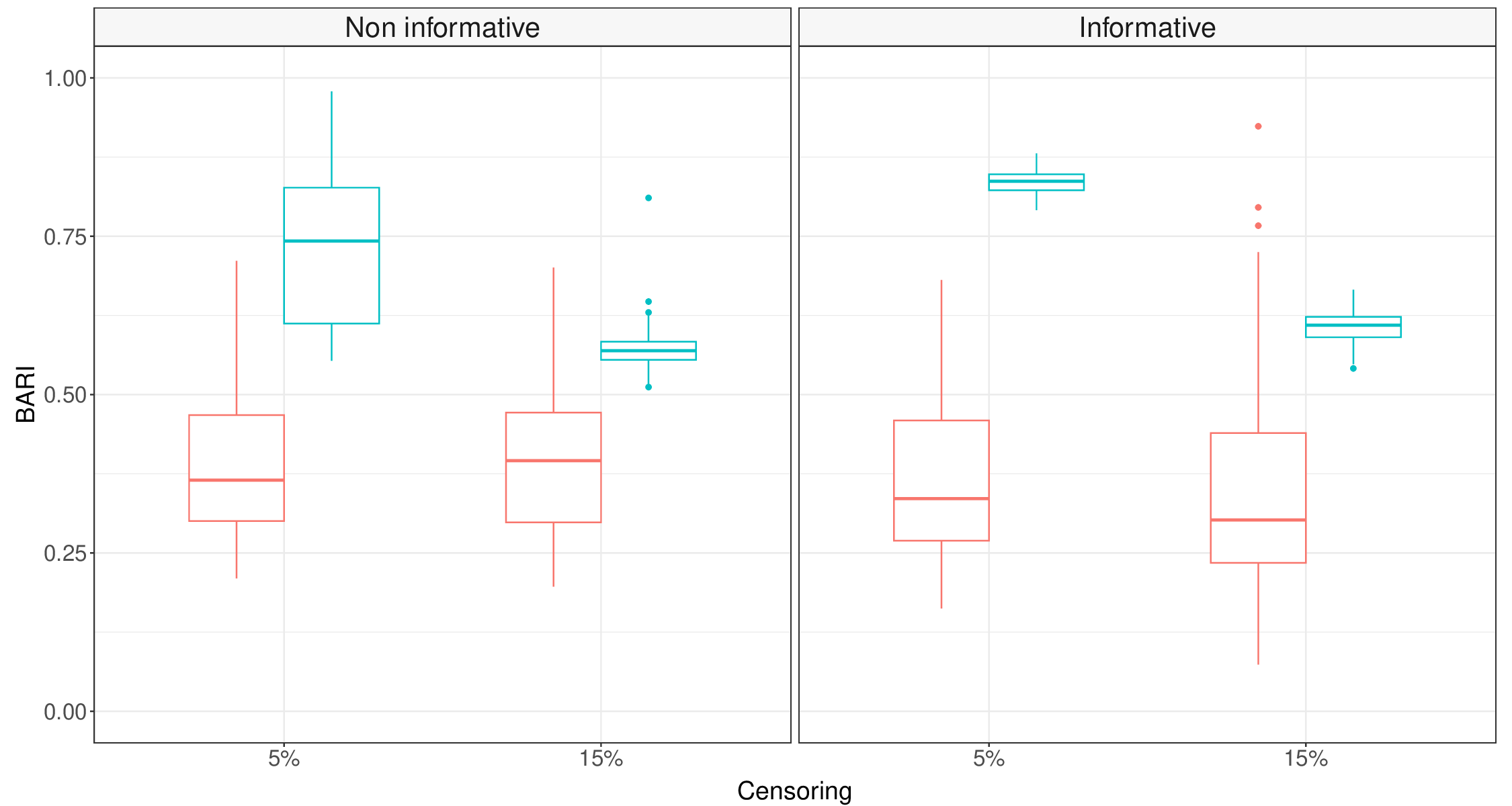}
    \caption{Simulated binary data. Boxplots for the BARI comparing the true bivariate partition and those identified by \texttt{biclustermd} (left boxplots) and \model (right boxplots). This comparison is made for datasets with 5\% or 15\% of entries missing, generated by a non-informative censoring mechanism (left panel) or an informative one (right panel).}\label{fig:BRI_sc2_joint}
\end{figure}

We now turn to the second part of the simulation study, which investigates how \model performs when analyzing datasets characterized by different types of censoring mechanisms, and how it compares to \texttt{biclustermd}. In this experiment, the data dimensions are fixed at $n = p = 50$. The data are simulated to mimic the voting patterns of politicians, which will be discussed in Section \ref{senator}. In this context, we assume that observations $y_{ij}$ take values in \{0, 1\}, representing votes \{\text{no}, \text{yes}\} on a specific political query. We reproduce a scenario with three major political parties and three types of voting patterns, achieved by simulating $\btheta_{1i}$ and $\btheta_{2j}$ from three-component mixture models. A portion of the observations, specifically 5\% or 15\%, is censored under two different schemes: (i) randomly, referred to as \say{non-informative censoring}, and (ii) uniformly at random among the entries equal to 0, termed \say{informative censoring}. The latter simulates a mechanism where missing entries may indicate opposition to a political motion. For each scenario, 100 independent datasets are generated. Examples of datasets simulated under this framework are illustrated in Figures \ref{fig:info5} in the \SMnsp. Also for this second experiment, the latent dimension $d$ was determined using the LPML criterion. The right panel of Figure \ref{fig:lpml_joint} indicates that $d = 3$ is the optimal choice in this scenario as well.\\
The results of the experiment, displayed in Figure \ref{fig:BRI_sc2_joint}, confirm the findings from the first part of the study, indicating that \model outperforms \texttt{biclustermd} across all scenarios, with consistently larger values for the BARI. Notably, \model excels in scenarios involving informative censoring, demonstrating its ability to exploit this additional source of information. Nevertheless, its performance remains robust even in scenarios where the censoring mechanism characterizing the data-generating process is non-informative. Overall, and as expected, the true latent clusters are identified more accurately in settings with 5\% censorship compared to those with 15\%.

\section{Real Data Illustrations}\label{sec:datareal}
We demonstrate the functionality of \model by analyzing two real-world datasets. The first dataset contains votes from U.S. senators and is characterized by binary responses, as discussed in one of the simulation studies in Section \ref{sec:simch2}. The second dataset relates to movie rankings, which are characterized by ordinal responses, as explored in the other simulation study in Section \ref{sec:simch2}. For both analyses, the hyperparameters are specified as outlined in Section \ref{sec:simch2}, with the exception of the DPs' concentration parameters, which are set equal to $\alpha_1 = 10^{-5}$ and $\alpha_2 = 10^{5}$, respectively, to favor robust inference.

\subsection{U.S. Senators Data}\label{senator}
Political data offer valuable insights into voting behavior within parties, revealing, for example, whether politicians consistently follow party lines or show divergent voting patterns. The dataset we consider was retrieved from voteview.com \citep{voteview} and it includes the voting records of 100 U.S. senators across 35 voting sessions held between May 2, 2022, and May 16, 2022. Notably, 3.37\% of the data entries are missing, corresponding to instances where a senator did not participate in a given session. It is reasonable to assume that censored observations hold valuable information, given that the choice to abstain from voting in a particular session is frequently a political statement in its own right. As for the simulated data considered in Section \ref{sec:simch2}, we resort to the LPML to set the value of the latent dimension $d$. Our analysis suggests that the best predictive performance is achieved when setting $d=3$. See Figure \ref{fig:lpml_sen} in the \SMnsp for details. 

\begin{figure}
    \centering
    \includegraphics[height=6cm]{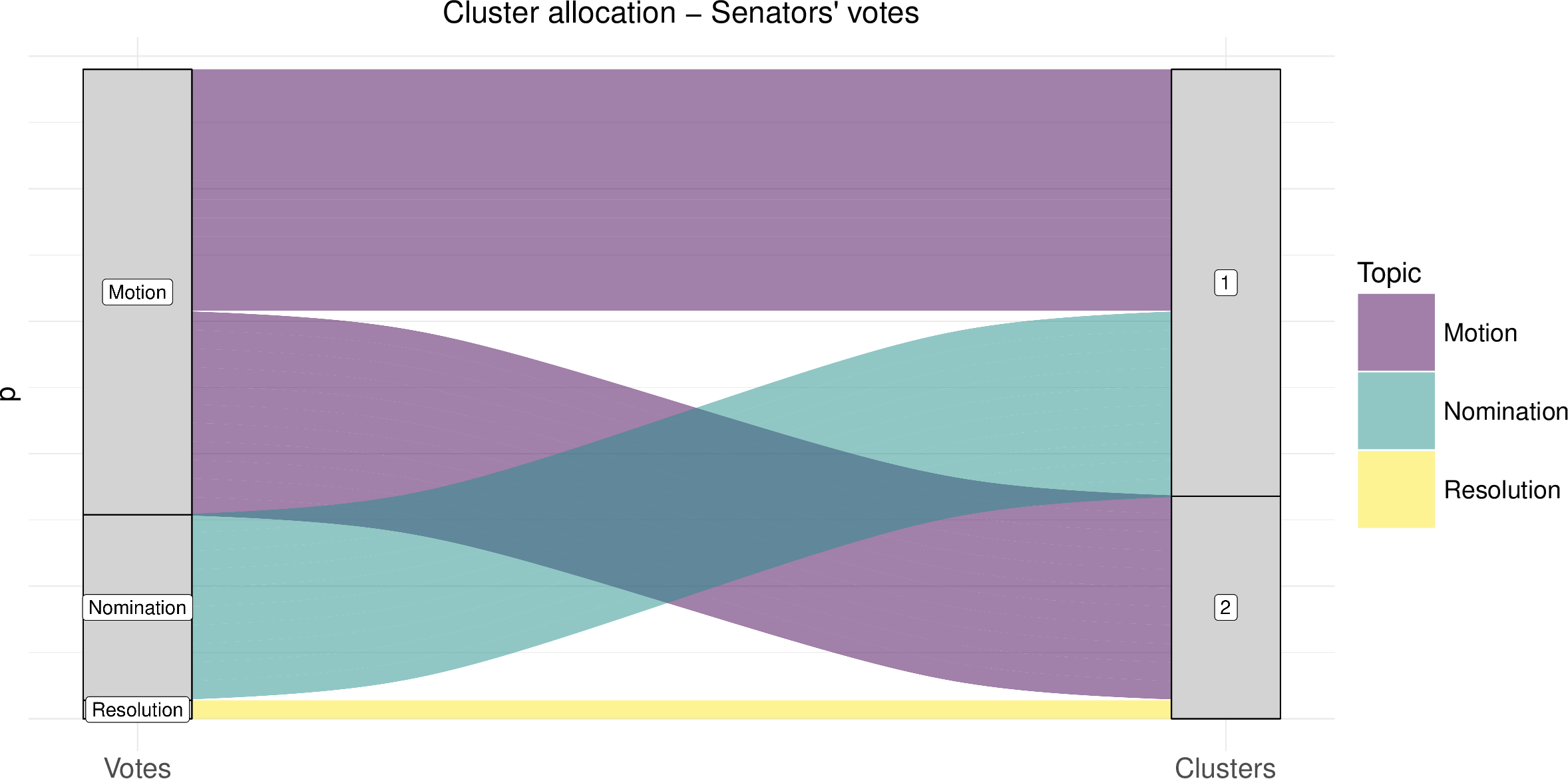}
    \includegraphics[height=6cm]{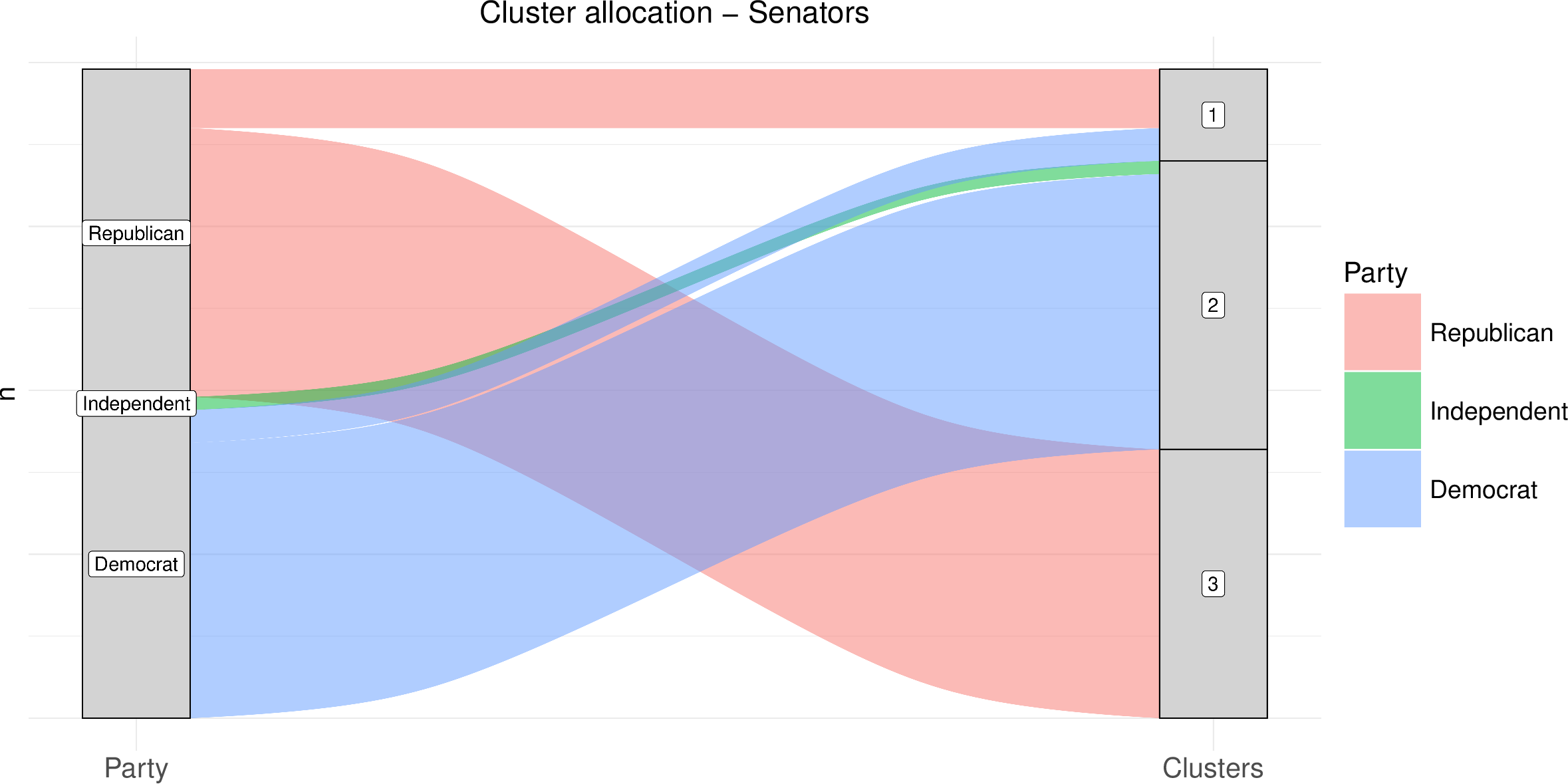}
    \caption{U.S. Senators Data. Alluvial diagrams comparing the estimated marginal clusters of votes (upper panel) and senators (lower panel). On the left of the plots, we report the topic of voting sessions (upper panel) and party affiliation (lower panel).}
    \label{fig:col_clu_sen}
\end{figure}

The results of our analysis, summarized by the alluvial diagrams in Figure \ref{fig:col_clu_sen}, offer valuable insights by comparing the identified clusters with available information on voting session types and party affiliations. For instance, voting sessions can be categorized into nominations to appoint someone to a specific position, motions, and resolutions. Our analysis identifies one large cluster containing nearly all the nominations and just over half of the motions, and a smaller cluster dominated by motions and resolutions. This outcome suggests potential similarities in the way senators voted on particular motions and nominations. Specifically, the cluster labeled 1 in the top panel of Figure \ref{fig:col_clu_sen} mainly consists of motions and nominations related to appointing individuals to specific roles, while cluster 2 in the same plot is primarily composed of motions on bills supporting federal research initiatives to maintain U.S. leadership in engineering biology and resolutions addressing various health policy changes. At the same time, examining the marginal results for the Senators, shown in the bottom panel of Figure \ref{fig:col_clu_sen}, reveals a significant polarization along party lines. Notably, independent senators are grouped with their Democratic counterparts. Additionally, a smaller third cluster indicates that the votes of nine Republican senators align with those of five Democratic senators. The names of these senators are listed in Table \ref{table:cluster3pol} in the \SMnsp: the composition of this third cluster may offer valuable insights for analysts studying U.S. parliamentary dynamics. 

\subsection{Movielens Data}\label{movie}

\begin{figure}[t]
    \centering
    \includegraphics[scale=0.3]{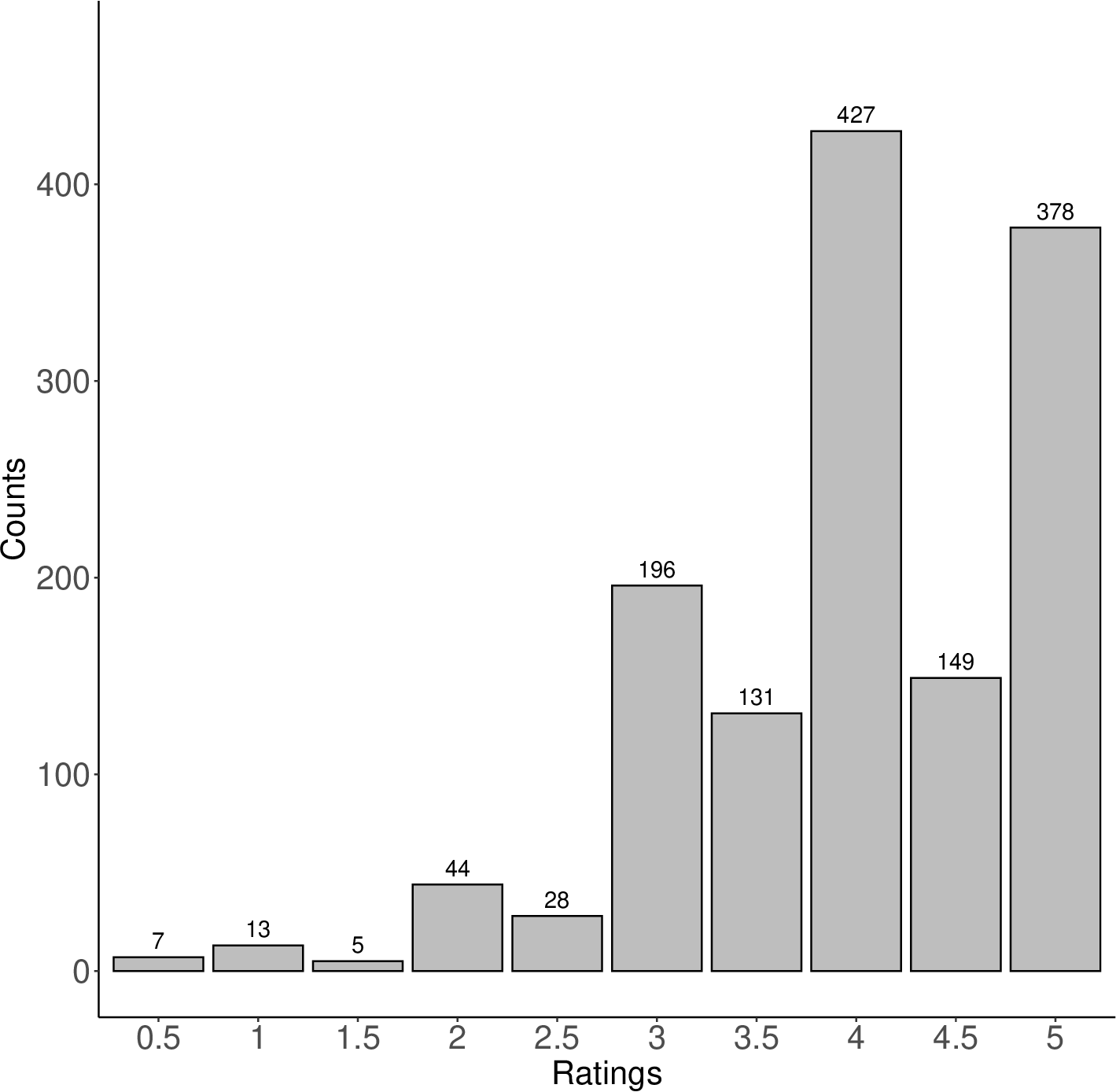}
    \caption{Movielens Data. Empirical frequencies of the movie ratings.}\label{fig:ratings}
\end{figure}

As a second illustrative application, we analyze a portion of the Movielens dataset, available in the \texttt{R} package \texttt{dslabs} \citep{dslabs}, which includes data from 60 users and 28 movies, and for which $17.98\%$ of the entries are missing. The ratings range from $\{0.5, 1, \ldots, 5\}$, with Figure \ref{fig:ratings} displaying the corresponding frequencies. The distribution of ratings is heavily concentrated around values of 3 and above, likely due to the dataset consisting of well-known and critically acclaimed movies. Additionally, the frequencies for half-point ratings, such as $\{0.5, 1.5, 2.5, 3.5, 4.5\}$, are lower than those for whole-number ratings, which include $\{1, 2, 3, 4, 5\}$. 
As with the other datasets in this work, the latent dimension $d$ was selected by running the analysis with different values of $d$ and evaluating the corresponding LPML. This analysis indicated that $d=2$ is optimal, as shown in Figure \ref{fig:lpml_movie} in the \SMnsp.

The results reveal four distinct groups of movies, henceforth labeled as clusters 1, 2, 3, and 4 for convenience. A closer examination of these groups shows that they are well differentiated by the genres of the movies they contain. Cluster 1 represents the genre \say{Drama/Thriller}, with the only surprising inclusion being \say{Toy Story}, which, based on its tags, seems an unexpected fit for this cluster. Cluster 2 comprises \say{Adventure/Action} movies and exhibits a rather homogeneous composition. The movies in Cluster 3 are characterized by plots involving a journey that the characters undertake to resolve their misadventures. In contrast, Cluster 4, which also appears rather homogeneous, consists of more satirical movies. Table \ref{tab:title} lists the titles of the 28 movies in the dataset, along with their genres and cluster allocations.

\begin{table}[h]
    \centering
    \adjustbox{max width=0.8\textwidth}{%
    \begin{tabular}{|c|c|c|}
    \hline
    \text{Title} & \text{Genre} & \text{Cluster} \\
    \hline
    ``Seven'' & Mystery|Thriller & 1 \\
    ``Pulp Fiction'' & Comedy|Crime|Drama|Thriller & 1 \\
    ``The Silence of the Lambs'' & Crime|Horror|Thriller & 1 \\
    ``The Shawshank Redemption'' & Crime|Drama & 1 \\
    ``The Sixth Sense'' & Drama|Horror|Mystery & 1 \\
    ``American Beauty'' & Drama|Romance & 1 \\
    ``The Godfather'' & Crime|Drama & 1 \\
    ``The Matrix'' & Action|Sci-Fi|Thriller & 1 \\
    ``Toy Story'' & Adventure|Animation|Children|Comedy|Fantasy & 1 \\
    \hline
    ``Braveheart'' & Action|Drama|War & 2 \\
    ``Forrest Gump'' & Comedy|Drama|Romance|War & 2 \\
    ``Speed'' & Action|Romance|Thriller & 2 \\
    ``Jurassic Park'' & Action|Adventure|Sci-Fi|Thriller & 2 \\
    ``Star Wars: Ep. VI'' & Action|Adventure|Sci-Fi & 2 \\
    ``Men in Black'' & Action|Comedy|Sci-Fi & 2 \\
    ``Star Wars: Ep. IV'' & Action|Adventure|Sci-Fi & 2 \\
    ``Die Hard'' & Action|Crime|Thriller & 2 \\
    ``Star Wars: Ep. V'' & Action|Adventure|Sci-Fi & 2 \\
    ``Raiders of the Lost Ark'' & Action|Adventure & 2 \\
    ``The Terminator'' & Action|Sci-Fi|Thriller & 2 \\
    ``Back to the Future'' & Adventure|Comedy|Sci-Fi & 2 \\
    \hline
    ``The Fugitive'' & Thriller & 3 \\
    ``E.T. the Extra-Terrestrial'' & Children|Drama|Sci-Fi & 3 \\
    ``Groundhog Day'' & Comedy|Fantasy|Romance & 3 \\
    ``Ferris Bueller's Day Off'' & Comedy & 3 \\
    \hline
    ``Monty Python and the Holy Grail'' & Adventure|Comedy|Fantasy & 4 \\
    ``Goodfellas'' & Crime|Drama & 4 \\
    ``Fargo'' & Comedy|Crime|Drama|Thriller & 4 \\
    \hline
    \end{tabular}
    }
    \caption{Movielens Data. Titles, genres, and estimated cluster allocations for the 28 movies in the dataset.} \label{tab:title}
\end{table}

The estimated partition of the 60 users in the dataset reveals six clusters with frequencies of $\{8, 31, 4, 7, 9, 1\}$, which we will refer to as Clusters 1, 2, 3, 4, 5, and 6 for convenience. To protect user privacy, no individual information was provided in the original dataset. Therefore, to gain insights into the composition of each user cluster, we analyze how users in each group rated movies across the four movie clusters: \say{Drama/Thriller}, \say{Adventure/Action}, \say{Journey}, and \say{Satire}.

\begin{figure}[h!]
    \centering
    \includegraphics[scale=0.7]{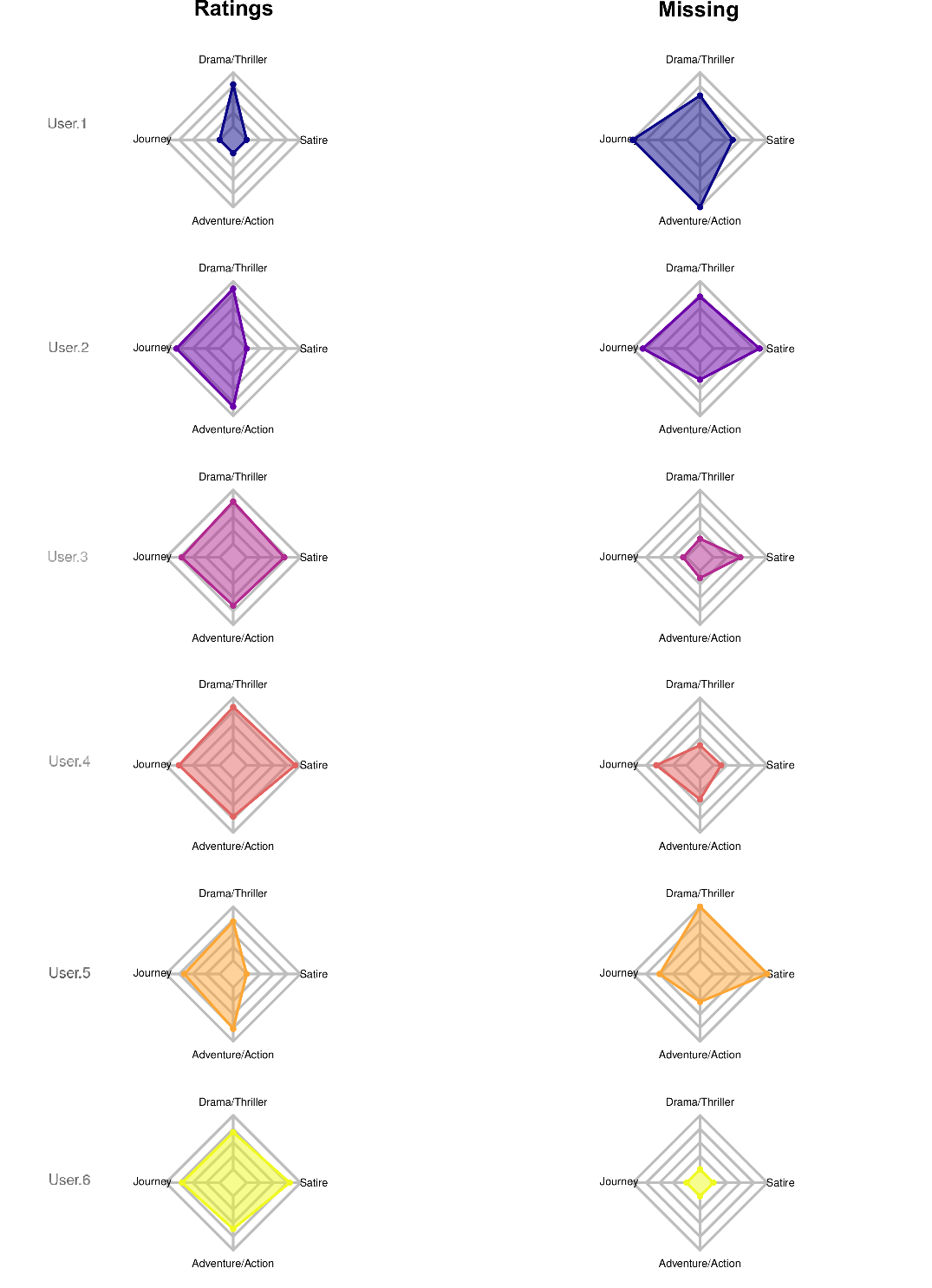}
    \caption{Movielens Data. Radar charts depicting the characterization of user clusters based on the main genres of the movies in each cluster. Ratings are displayed on the left, the percentage of missing values is shown on the right.}
    \label{fig:radar}
\end{figure}

Examining Figure~\ref{fig:radar}, it is clear that different user clusters exhibit varying degrees of appreciation for the four identified movie genres. It is important to note that while high ratings certainly reflect a positive view of a movie, missing ratings may indicate a lack of interest. In the left column of Figure \ref{fig:radar}, we observe the rating patterns of users across the different clusters. Clusters 2 and 5 show similar rating patterns, as do Clusters 3 and 6. However, the right column, which displays the percentage of missing values for each user cluster, reveals distinct differences in the missing data. This visualization highlights the significance of considering missing entries as valuable information.

\section{Discussion}

We introduced \modelnosp, a nonparametric Bayesian method for co-clustering the rows and columns of a matrix of ordinal data, accommodating potentially informative missing entries. Our method employs matrix factorization to reduce the problem’s dimensionality, while the ordinal nature of the data is handled by introducing continuous latent variables, which facilitates model implementation. \model fills a gap in the literature, as no model-based approach for co-clustering ordinal data with missing entries had previously been introduced. Consequently, in our simulation study, we compared \model with a geometric approach for co-clustering that accounts for missing data, implemented in the \texttt{R} package \texttt{biclustermd}. The results consistently demonstrated the superior performance of \model over the geometric method. Notably, \model proved robust even with synthetic data featuring randomly censored, and thus non-informative, missing entries. Moreover, when applied to the U.S. Senators and Movielens Data, \model produced interpretable and valuable results for domain experts.

The definition of \model relies on two independent DPs to co-cluster the rows and columns of the data matrix. Posterior simulation is achieved via a Gibbs sampling algorithm with two generalized P\'olya urn steps. This structure makes \model highly flexible and able to infer the number of clusters in rows and columns from the data. However, this flexibility also makes posterior inference somewhat sensitive to model hyperparameters, which is a well-known limitation of models with this level of adaptability. In addition, it should be considered that Algorithm \ref{gibbs} does not scale well when $n$ and $p$ are large, primarily due to the use of a marginal algorithm, resulting from the analytical marginalization of the DPs, and the factorization applied to model the mean of continuous latent variables in \eqref{eq:modelgeneral}. Addressing these limitations is essential to extend similar modeling strategies to large datasets like the Netflix Prize data\footnote{\url{https://www.kaggle.com/datasets/netflix-inc/netflix-prize-data}}. This may necessitate a different computational approach, with one promising option being an adaptation of the scalable multistep Monte Carlo algorithm by \cite{ni2020scalable}. Moreover, handling extensive missing data—as is typical in the Netflix Prize data—may require an alternative modeling approach tailored to such scenarios.

\clearpage
\bibliographystyle{chicago}
\bibliography{main}

\clearpage
\renewcommand{\thefigure}{\thesection\arabic{figure}}
\renewcommand{\thetable}{\thesection\arabic{table}}
\setcounter{table}{0}
\setcounter{figure}{0}

\section*{Supplementary Material}

This file is organized as follows. In Section \ref{factormodel} we report additional information on the full conditional distributions involved in the Gibbs sampling algorithm summarized in Algorithm \ref{gibbs}. Section \ref{app:sim} and \ref{app:real} provide additional results on the simulation study of Section \ref{sec:simch2} and the data analyses of Section \ref{sec:datareal}, respectively, 

\appendix

\section{Full conditional distributions}\label{factormodel}
We provide additional details on the full conditional distributions already discussed in Section \ref{sec:postch2}. To this end, introducing additional notation is going to be useful. For $r=1,2$, we denote the $r$-th column of $\bm{M}_l$ as $\bm{m}_{lr}=(m_{lr1},\ldots,m_{lrd})$; moreover, we let $v_{ls_1s_2}$ denote the element in position $(s_1,s_2)$ of  matrix $\bm{V}_l$ and $u_{lr_1r_2}$ denote the element in position $(r_1,r_2)$ of  matrix $\bm{U}_l$.\\

\noindent The weights in \eqref{eq:fc_theta_1} are given, up to a proportionality constant, by
\begin{align*}
\pi_{1i0}&\propto \frac{\alpha_1}{\alpha_1+n-1}
|\tilde{\bm{V}_1}|^{-\frac{1}{2}} \frac{1}{\sqrt{u_{122}}} |\bm{V}_1|^{-\frac{1}{2}} |(\frac{1}{\sigma_1^2}\btheta_2^{(1)\intercal}\btheta_2^{(1)} + \tilde{\bm{V}}_1^{-1})|^{-\frac{1}{2}}\\
&\times |(\frac{1}{\sigma_2^2}\btheta_2^{(2)\intercal}\btheta_2^{(2)} + \frac{1}{u_{122}}{\bm{V}}_1^{-1})|^{-\frac{1}{2}} \exp \left\{-\frac{1}{2} \tr\left[\frac{1}{\sigma_1^2}\bm{z}_i\bm{z}_i^\intercal + \frac{1}{\sigma_2^2}\bm{w}_i\bm{w}_i^\intercal \right]\right\}\\
&\times \exp \left\{-\frac{1}{2} \tr\left[ \tilde{\bm{V}}_1^{-1}\tilde{\bm{m}}_{11}\tilde{\bm{m}}_{11}^\intercal+\frac{1}{u_{122}}\bm{V}_1^{-1} \bm{m}_{12} \bm{m}_{12}\right]\right\}\\
&\times \exp \left\{+\frac{1}{2} \tr\left[ 
\left(\frac{1}{\sigma_1^2} \btheta_2^{(1)\intercal} \bm{z}_i + \tilde{\bm{V}}_1^{-1} \tilde{\bm{m}}_{11}\right)\left(\frac{1}{\sigma_1^2} \btheta_2^{(1)\intercal} \bm{z}_i + \tilde{\bm{V}}_1^{-1} \tilde{\bm{m}}_{11}\right)^\intercal \left(\frac{1}{\sigma_1^2}\btheta_2^{(1)\intercal}\btheta_2^{(1)} + \tilde{\bm{V}}_1^{-1}\right)^{-1}
\right]\right\}\\
&\times \exp \left\{+\frac{1}{2} \tr\left[ 
\left(\frac{1}{\sigma_2^2} \btheta_2^{(2)\intercal} \bm{w}_i + \frac{1}{u_{122}}{\bm{V}}_1^{-1} {\bm{m}}_{12}\right)\left(\frac{1}{\sigma_2^2} \btheta_2^{(2)\intercal} \bm{w}_i + \frac{1}{u_{122}}{\bm{V}}_1^{-1} {\bm{m}}_{12}\right)^\intercal \left(\frac{1}{\sigma_2^2}\btheta_2^{(2)\intercal}\btheta_2^{(2)} + \frac{1}{u_{122}}{\bm{V}}_1^{-1}\right)^{-1}
\right]\right\}\\[6pt]
\pi&_{2i0}\propto \frac{\alpha_2}{\alpha_2+p-1}
|\tilde{\bm{V}_2}|^{-\frac{1}{2}} \frac{1}{\sqrt{u_{222}}} |\bm{V}_2|^{-\frac{1}{2}} |(\frac{1}{\sigma_1^2}\btheta_1^{(1)\intercal}\btheta_1^{(1)} + \tilde{\bm{V}}_2^{-1})|^{-\frac{1}{2}}
|(\frac{1}{\sigma_2^2}\btheta_1^{(2)\intercal}\btheta_1^{(2)} + \frac{1}{u_{222}}{\bm{V}}_2^{-1})|^{-\frac{1}{2}}\\
&\times \exp \left\{-\frac{1}{2} \tr\left[\frac{1}{\sigma_1^2}\bm{z}_j\bm{z}_j^\intercal + \frac{1}{\sigma_2^2}\bm{w}_j\bm{w}_j^\intercal \right]\right\}\\
&\times \exp \left\{-\frac{1}{2} \tr\left[ \tilde{\bm{V}}_2\tilde{\bm{m}}_{21}\tilde{\bm{m}}_{21}^\intercal+\frac{1}{u_{222}}\bm{V}_2^{-1} \bm{m}_{22} \bm{m}_{22}\right]\right\}\\
&\times \exp \left\{+\frac{1}{2} \tr\left[ 
\left(\frac{1}{\sigma_1^2} \btheta_1^{(1)\intercal} \bm{z}_j + \tilde{\bm{V}}_2^{-1} \tilde{\bm{m}}_{21}\right)\left(\frac{1}{\sigma_1^2} \btheta_1^{(1)\intercal} \bm{z}_j + \tilde{\bm{V}}_2^{-1} \tilde{\bm{m}}_{21}\right)^\intercal \left(\frac{1}{\sigma_1^2}\btheta_1^{(1)\intercal}\btheta_1^{(1)} + \tilde{\bm{V}}_2^{-1}\right)^{-1}
\right]\right\}\\
&\times \exp \left\{+\frac{1}{2} \tr\left[ 
\left(\frac{1}{\sigma_2^2} \btheta_1^{(2)\intercal} \bm{w}_j + \frac{1}{u_{222}}{\bm{V}}_2^{-1} {\bm{m}}_{22}\right)\left(\frac{1}{\sigma_2^2} \btheta_1^{(2)\intercal} \bm{w}_j + \frac{1}{u_{222}}{\bm{V}}_2^{-1} {\bm{m}}_{22}\right)^\intercal \left(\frac{1}{\sigma_2^2}\btheta_1^{(2)\intercal}\btheta_1^{(2)} + \frac{1}{u_{222}}{\bm{V}}_2^{-1}\right)^{-1}
\right]\right\}.
\end{align*}
Sampling from $G_{1i}$ in \eqref{eq:fc_theta_1} is straightforward after observing that $G_{1i}$ is the distribution of a $(d\times 2)$-dimensional matrix $\bm{T}$, with columns $\bm{t}_1$ and $\bm{t}_2$, for which
\begin{align*}
    \bm{t}_1\mid \bm{t}_2 &\sim \textsc{N}_d \left(\tilde{\bm{m}}_{11}, \tilde{\bm{V}_1} \right)\\
    \bm{t}_2 &\sim \textsc{N}_d \left(\bm{m}_{12}, u_{122} \bm{V}_1 \right).
\end{align*}
Similarly, the distribution $G_{2j}$ in \eqref{eq:fc_theta_2} coincides with the distributions of a $(d\times 2)$-dimensional matrix $\bm{T}$, with columns $\bm{t}_1$ and $\bm{t}_2$, for which
\begin{align*}
    \bm{t}_1\mid \bm{t}_2 &\sim \textsc{N}_d \left(\tilde{\bm{m}}_{21}, \tilde{\bm{V}_2} \right)\\
    \bm{t}_2 &\sim \textsc{N}_d \left(\bm{m}_{22}, u_{222} \bm{V}_2 \right).
\end{align*}

Next, we show how the weights in \eqref{eq:fc_theta_1} and \eqref{eq:fc_theta_2} simplify when the base measures $H_1$ and $H_2$ have independent components, which is achieved, respectively, when $u_{112}=u_{121}=0$ and $u_{212}=u_{221}=0$. 

\begin{align*}
\pi&_{1i0}\propto \frac{\alpha_1}{\alpha_1+n-1}
\frac{1}{\sqrt{u_{111}}}|{\bm{V}_1}|^{-\frac{1}{2}} \frac{1}{\sqrt{u_{122}}} |\bm{V}_1|^{-\frac{1}{2}} |(\frac{1}{\sigma_1^2}\btheta_2^{(1)\intercal}\btheta_2^{(1)} + \frac{1}{u_{111}}{\bm{V}}_1^{-1})|^{-\frac{1}{2}}
|(\frac{1}{\sigma_2^2}\btheta_2^{(2)\intercal}\btheta_2^{(2)} + \frac{1}{u_{122}}{\bm{V}}_1^{-1})|^{-\frac{1}{2}}\\
&\times \exp \left\{-\frac{1}{2} \tr\left[\frac{1}{\sigma_1^2}\bm{z}_i\bm{z}_i^\intercal + \frac{1}{\sigma_2^2}\bm{w}_i\bm{w}_i^\intercal \right]\right\}\\
&\times \exp \left\{-\frac{1}{2} \tr\left[ \frac{1}{u_{111}}{\bm{V}}_1^{-1}{\bm{m}}_{11}{\bm{m}}_{11}^\intercal+\frac{1}{u_{122}}\bm{V}_1^{-1} \bm{m}_{12} \bm{m}_{12}\right]\right\}\\
&\times \exp \left\{+\frac{1}{2} \tr\left[ 
\left(\frac{1}{\sigma_1^2} \btheta_2^{(1)\intercal} \bm{z}_i + \frac{1}{u_{111}}{\bm{V}}_1^{-1} {\bm{m}}_{11}\right)\left(\frac{1}{\sigma_1^2} \btheta_2^{(1)\intercal} \bm{z}_i + \frac{1}{u_{111}}{\bm{V}}_1^{-1} {\bm{m}}_{11}\right)^\intercal \left(\frac{1}{\sigma_1^2}\btheta_2^{(1)\intercal}\btheta_2^{(1)} + \frac{1}{u_{111}}{\bm{V}}_1^{-1}\right)^{-1}
\right]\right\}\\
&\times \exp \left\{+\frac{1}{2} \tr\left[ 
\left(\frac{1}{\sigma_2^2} \btheta_2^{(2)\intercal} \bm{w}_i + \frac{1}{u_{122}}{\bm{V}}_1^{-1} {\bm{m}}_{12}\right)\left(\frac{1}{\sigma_2^2} \btheta_2^{(2)\intercal} \bm{w}_i + \frac{1}{u_{122}}{\bm{V}}_1^{-1} {\bm{m}}_{12}\right)^\intercal \left(\frac{1}{\sigma_2^2}\btheta_2^{(2)\intercal}\btheta_2^{(2)} + \frac{1}{u_{122}}{\bm{V}}_1^{-1}\right)^{-1}
\right]\right\}\\[6pt]
\pi&_{2i0}\propto \frac{\alpha_2}{\alpha_2+p-1}
\frac{1}{\sqrt{u_{211}}}|{\bm{V}_2}|^{-\frac{1}{2}} \frac{1}{\sqrt{u_{222}}} |\bm{V}_2|^{-\frac{1}{2}} |(\frac{1}{\sigma_1^2}\btheta_1^{(1)\intercal}\btheta_1^{(1)} + \frac{1}{u_{211}}{\bm{V}}_2^{-1})|^{-\frac{1}{2}}
|(\frac{1}{\sigma_2^2}\btheta_1^{(2)\intercal}\btheta_1^{(2)} + \frac{1}{u_{222}}{\bm{V}}_2^{-1})|^{-\frac{1}{2}}\\
&\times \exp \left\{-\frac{1}{2} \tr\left[\frac{1}{\sigma_1^2}\bm{z}_j\bm{z}_j^\intercal + \frac{1}{\sigma_2^2}\bm{w}_j\bm{w}_j^\intercal \right]\right\}\\
&\times \exp \left\{-\frac{1}{2} \tr\left[ \frac{1}{u_{211}}{\bm{V}}_2^{-1}{\bm{m}}_{21}{\bm{m}}_{21}^\intercal+\frac{1}{u_{222}}\bm{V}_2^{-1} \bm{m}_{22} \bm{m}_{22}\right]\right\}\\
&\times \exp \left\{+\frac{1}{2} \tr\left[ 
\left(\frac{1}{\sigma_1^2} \btheta_1^{(1)\intercal} \bm{z}_j + \frac{1}{u_{211}}{\bm{V}}_2^{-1} {\bm{m}}_{21}\right)\left(\frac{1}{\sigma_1^2} \btheta_1^{(1)\intercal} \bm{z}_j + \frac{1}{u_{211}}{\bm{V}}_2^{-1} {\bm{m}}_{21}\right)^\intercal \left(\frac{1}{\sigma_1^2}\btheta_1^{(1)\intercal}\btheta_1^{(1)} + \frac{1}{u_{211}}{\bm{V}}_2^{-1}\right)^{-1}
\right]\right\}\\
&\times \exp \left\{+\frac{1}{2} \tr\left[ 
\left(\frac{1}{\sigma_2^2} \btheta_1^{(2)\intercal} \bm{w}_j + \frac{1}{u_{222}}{\bm{V}}_2^{-1} {\bm{m}}_{22}\right)\left(\frac{1}{\sigma_2^2} \btheta_1^{(2)\intercal} \bm{w}_j + \frac{1}{u_{222}}{\bm{V}}_2^{-1} {\bm{m}}_{22}\right)^\intercal \left(\frac{1}{\sigma_2^2}\btheta_1^{(2)\intercal}\btheta_1^{(2)} + \frac{1}{u_{222}}{\bm{V}}_2^{-1}\right)^{-1}
\right]\right\}.
\end{align*}
Also sampling from 
$G_{1i}$ and $G_{2j}$ simplifies. $G_{1i}$ is the distribution of a $(d\times 2)$-dimensional matrix $\bm{T}$, with independent columns $\bm{t}_1$ and $\bm{t}_2$ such that

\begin{align*}
    \bm{t}_1 &\sim \textsc{N}_d \left({\bm{m}}_{11}, u_{111}{\bm{V}_1} \right)\\
    \bm{t}_2 &\sim \textsc{N}_d \left(\bm{m}_{12}, u_{122} \bm{V}_1 \right).
\end{align*}
Similarly, for $\bm{T}\sim G_{2j}$ we have
\begin{align*}
    \bm{t}_1 &\sim \textsc{N}_d \left({\bm{m}}_{21}, u_{211}{\bm{V}_2} \right)\\
    \bm{t}_2 &\sim \textsc{N}_d \left(\bm{m}_{22}, u_{222} \bm{V}_2 \right).
\end{align*}

We conclude this section by discussing how to implement the reshuffling step in \eqref{eq:acc}. For any $\ell_1=1,\ldots,k_n$, we let $\bar{\bm{z}}_{\ell_1}$ and $\bar{\bm{w}}_{\ell_1}$ be, respectively, the component-wise averages of the vectors $\bm{z}_i$ and $\bm{w}_i$ for $i\in \mathcal{C}_{\ell_1}$.  
By considering again the simplifying assumption $u_{112} = u_{121} =0$, we obtain
\begin{align*}
    \btheta_{1\ell_1}^{(1)\ast}\mid \ldots \simind \text{N}_d\big((u_{111}^{-1}\bm{V}_1^{-1} +n_{\ell_1} \frac{1}{\sigma_1^2}\btheta_2^{(1)}\btheta_2^{(1)\intercal})^{-1}&(u_{111}^{-1}\bm{V}_1^{-1} \bm{m}_{11}+n_{\ell_1} \frac{1}{\sigma_1^2}\btheta_2^{(1)} \bar{\bm{z}}_{\ell_1}), \\&(u_{111}^{-1}\bm{V}_1^{-1} +n_{\ell_1} \frac{1}{\sigma_1^2}\btheta_2^{(1)}\btheta_2^{(1)\intercal})^{-1}\big)\\
   \btheta_{1\ell_1}^{(2)\ast}\mid\ldots \simind \text{N}_d\big((u_{122}^{-1}\bm{V}_1^{-1} +n_{\ell_1} \frac{1}{\sigma_2^2}\btheta_2^{(2)}\btheta_2^{(2)\intercal})^{-1}&(u_{122}^{-1}\bm{V}_1^{-1}\bm{m}_{12}+n_{\ell_1}\frac{1}{\sigma_2^2}\btheta_2^{(2)}\bar{\bm{w}}_{\ell_1}),\\& (u_{122}^{-1}\bm{V}_1^{-1} +n_{\ell_1} \frac{1}{\sigma_2^2}\btheta_2^{(2)}\btheta_2^{(2)\intercal})^{-1} \big)
\end{align*}

Similarly, for any $\ell_2=1,\ldots,k_p$, we let $\bar{\bm{z}}_{\ell_2}$ and $\bar{\bm{w}}_{\ell_2}$ be, respectively, the component-wise averages of the vectors $\bm{z}_j$ and $\bm{w}_j$ for $j\in \mathcal{C}_{\ell_2}$.
By considering the simplifying assumption $u_{212} = u_{221} =0$, we obtain
\begin{align*}
    \btheta_{2\ell_2}^{(1)\ast}\mid \ldots \simind \text{N}_d\big((u_{211}^{-1}\bm{V}_2^{-1} +n_{\ell_2} \frac{1}{\sigma_1^2}\btheta_1^{(1)}\btheta_1^{(1)\intercal})^{-1}&(u_{211}^{-1}\bm{V}_2^{-1} \bm{m}_{21}+n_{\ell_2} \frac{1}{\sigma_1^2}\btheta_1^{(1)} \bar{\bm{z}}_{\ell_2}), \\&(u_{211}^{-1}\bm{V}_2^{-1} +n_{\ell_2} \frac{1}{\sigma_1^2}\btheta_1^{(1)}\btheta_1^{(1)\intercal})^{-1}\big)\\
   \btheta_{2\ell_2}^{(2)\ast}\mid\ldots \simind \text{N}_d\big((u_{222}^{-1}\bm{V}_2^{-1} +n_{\ell_2} \frac{1}{\sigma_2^2}\btheta_1^{(2)}\btheta_1^{(2)\intercal})^{-1}&(u_{222}^{-1}\bm{V}_2^{-1}\bm{m}_{22}+n_{\ell_2}\frac{1}{\sigma_2^2}\btheta_1^{(2)}\bar{\bm{w}}_{\ell_2}),\\& (u_{222}^{-1}\bm{V}_2^{-1} +n_{\ell_2} \frac{1}{\sigma_2^2}\btheta_1^{(2)}\btheta_1^{(2)\intercal})^{-1} \big)
\end{align*}

\section{Additional details on the simulation study}\label{app:sim}

In this section, we provide additional details on the simulation study of Section \ref{sec:simch2}.

\begin{figure}[h!]
    \centering
    \includegraphics[clip,width=0.5\textwidth]{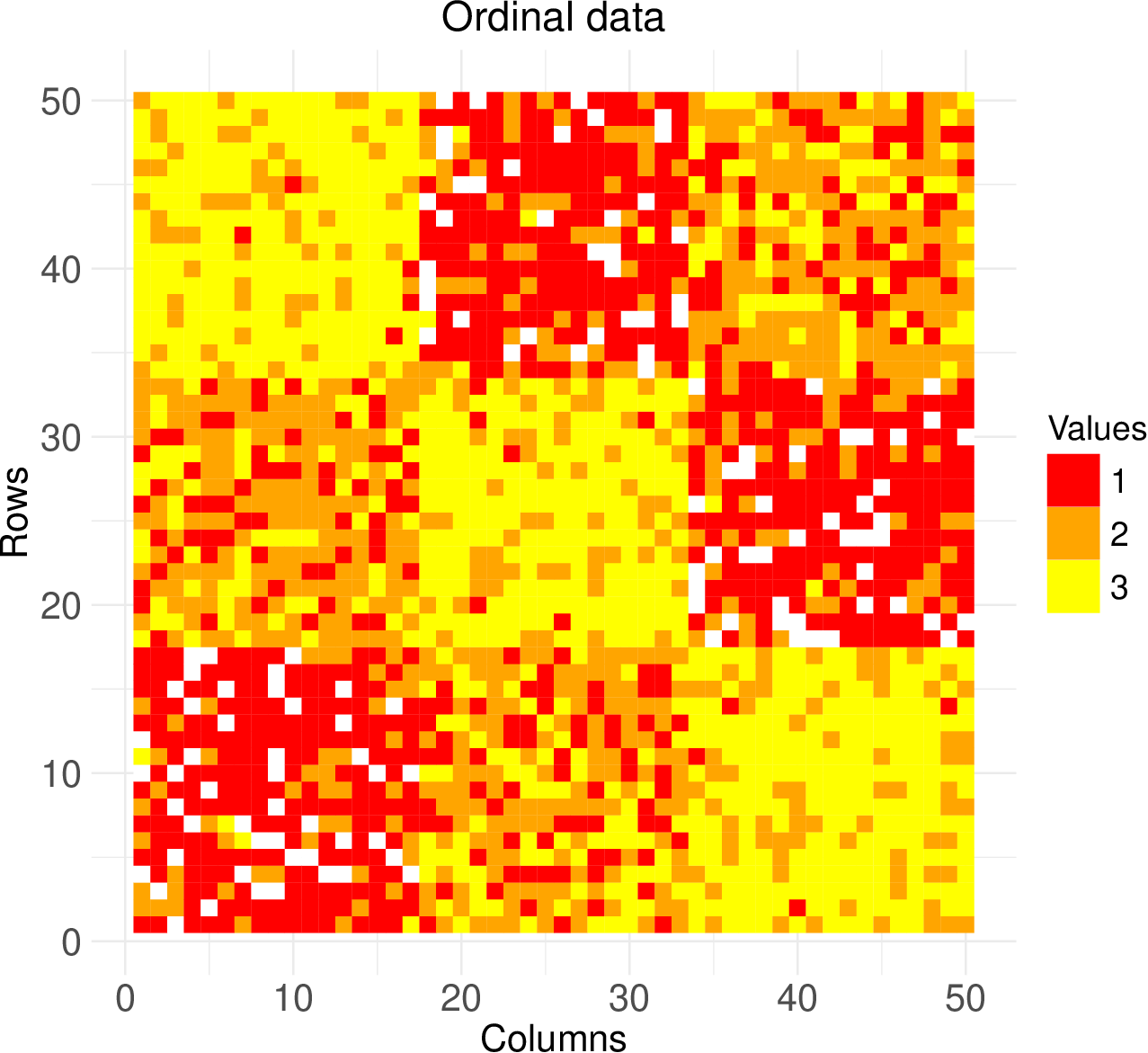}
    \caption{Graphical representation of a simulated dataset with ordinal observations. White pixels (5\% of the total) indicate missing entries generated according to an informative censoring mechanism. Refer to Section \ref{sec:simch2} for further details.\label{fig:ordinal}}
\end{figure}

\begin{table}[h!]
\centering
\caption{Simulated data with ordinal observations. Estimated ARI for the marginal partitions of the rows and columns of $\bm{Y}$, along with the estimated number of clusters, $\hat{k}_n$ and $\hat{k}_p$, for varying dataset sizes. Values in parentheses represent the estimated standard error. Refer to Section \ref{sec:simch2} for further details.} \label{tab:mari_sc1}
\begin{tabular}{ccccccc}
& & \multicolumn{2}{c}{Rows} & & \multicolumn{2}{c}{Columns} \\
\hline
$n\times p$ & & ARI & $\hat{k}_n$ & & ARI & $\hat{k}_p$\\
\hline
50$\times$50 & & 0.880 (0.059)& 2.006 (0.826) & & 0.932 (0.013) & 2.451 (0.973)\\
100$\times$100  & & 0.895 (0.065)& 3.257 (1.490) & & 0.961 (0.008)& 2.556 (0.998)\\
200$\times$200  & & 0.957 (0.019)& 3.067 (1.609)& & 0.970 (0.001)& 3.809 (1.103)\\
\hline
\end{tabular}
\end{table}

\begin{figure}[h!]
    \centering
    \includegraphics[clip,width=0.48\textwidth]{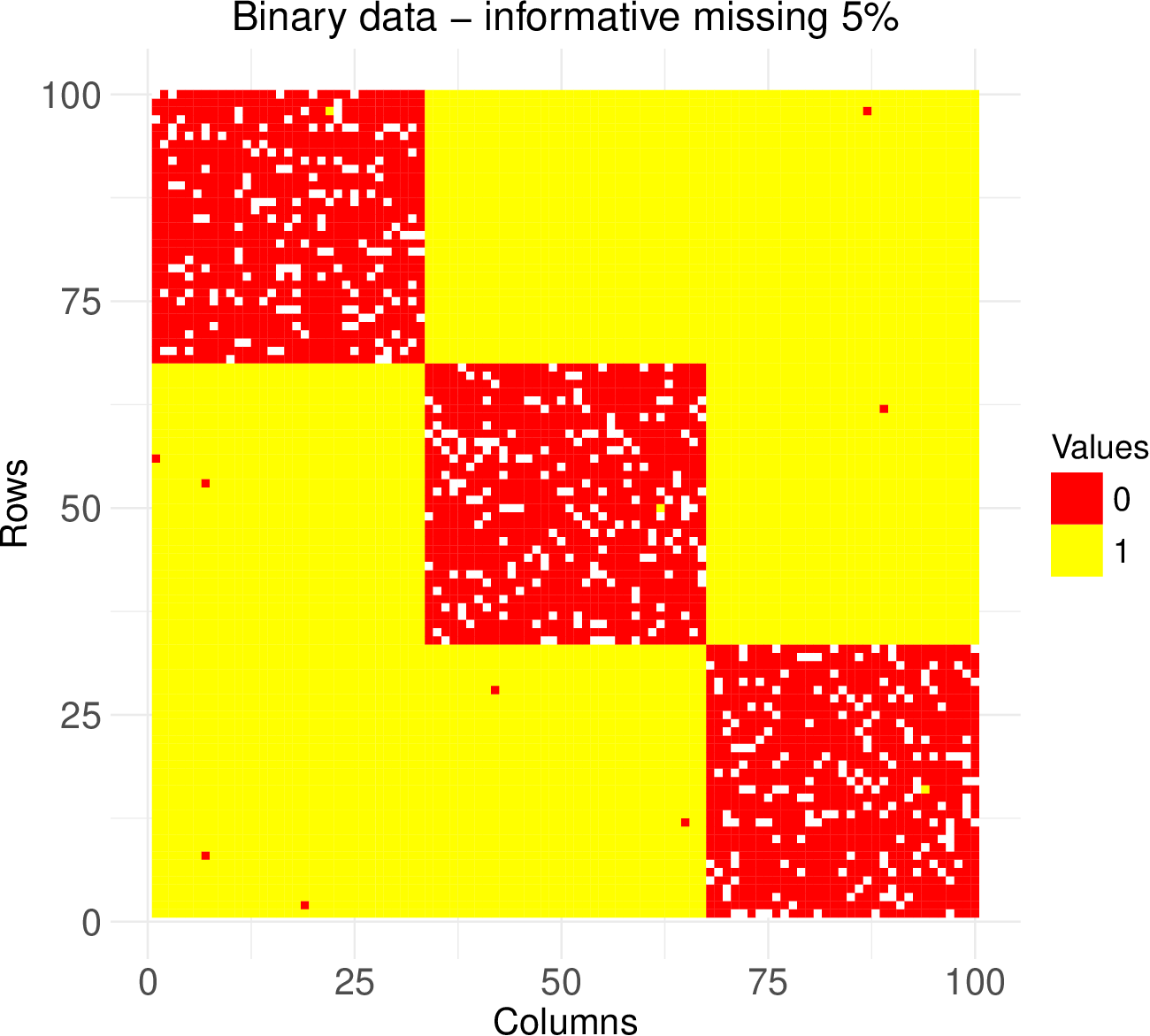}
    \includegraphics[clip,width=0.48\textwidth]{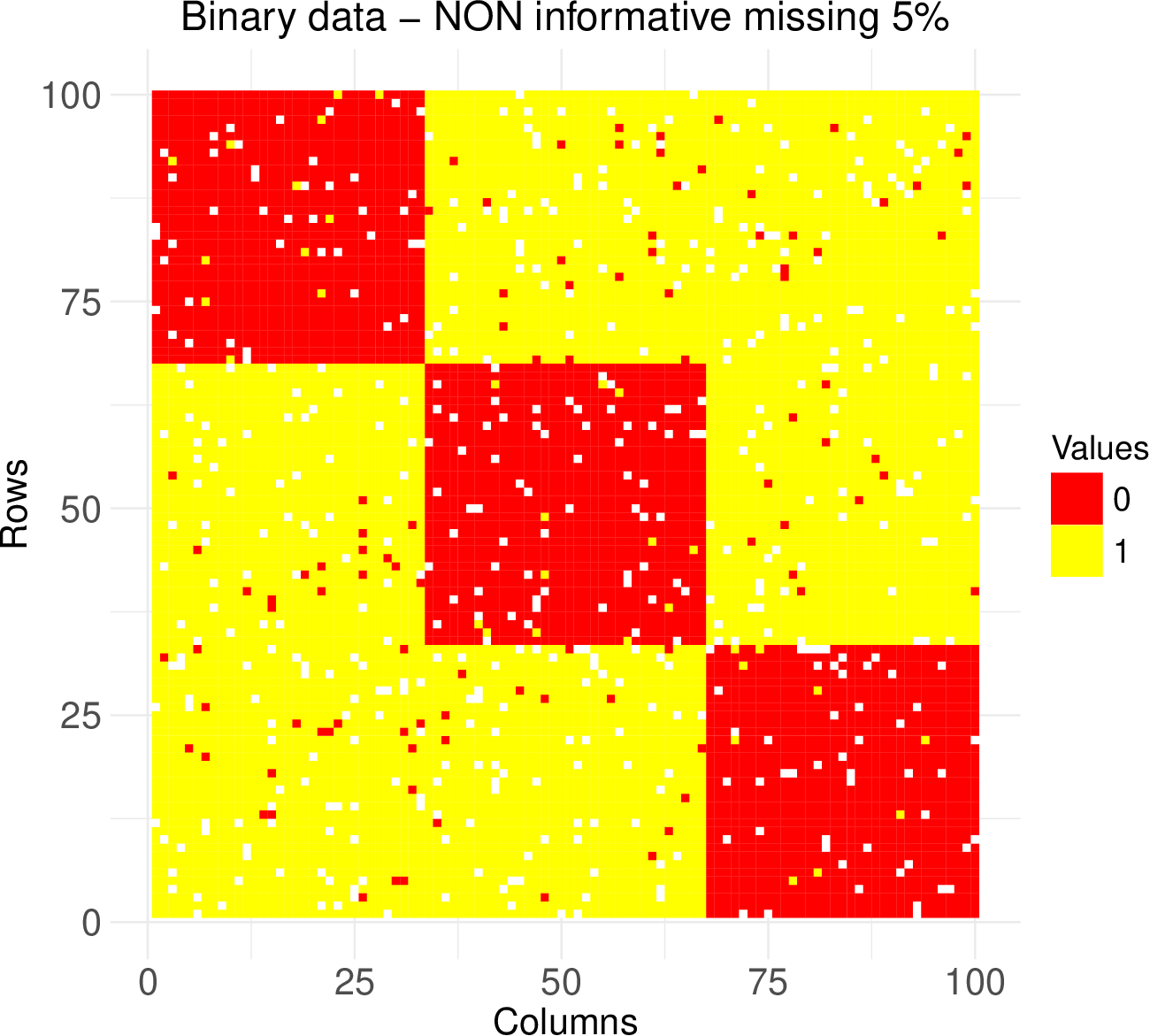}
    \caption{Graphical representation of a simulated dataset with binary observations. White pixels (5\% of the total) indicate missing entries generated according to an informative censoring mechanism (left panel) or to a non-informative censoring mechanism (right panel). Refer to Section \ref{sec:simch2} for further details.}\label{fig:info5}
\end{figure}

\section{Additional details on the real data analyses}\label{app:real}

In this section, we provide additional details of the analysis of datasets we presented in \ref{sec:datareal}.

\subsection{U.S. Senate data}

\begin{figure}[h!]
    \centering
    \includegraphics[height=8cm]{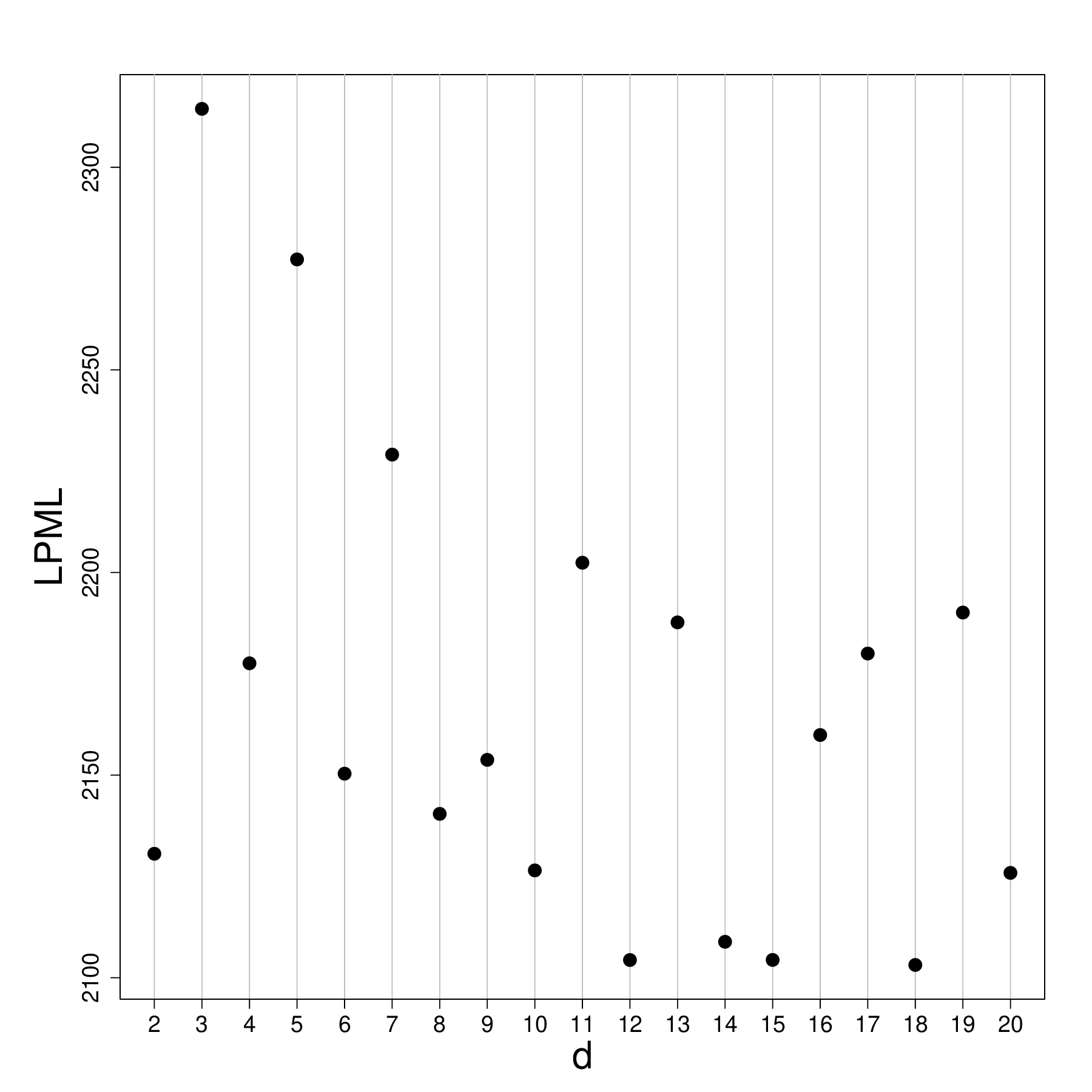}
    \caption{U.S. Senators Data. LPML as a function of the latent dimension $d$. Refer to Section \ref{senator} for further details.}
    \label{fig:lpml_sen}
\end{figure}

\begin{table}[ht]
\centering
\begin{tabular}{ll}
  \hline
  Name & Party \\ 
  \hline
 BLUMENTHAL, Richard & Democrat \\ 
   DURBIN, Richard Joseph & Democrat \\ 
   FEINSTEIN, Dianne & Democrat \\ 
   FISCHER, Debra (Deb) & Republican \\ 
   JOHNSON, Ron & Republican \\ 
   LEE, Mike & Republican \\ 
   MARKEY, Edward John & Democrat \\ 
   MORAN, Jerry & Republican \\ 
   PAUL, Rand & Republican \\ 
   RUBIO, Marco & Republican \\ 
   SCHATZ, Brian Emanuel & Democrat \\ 
  SCOTT, Tim & Republican \\ 
   SHELBY, Richard C. & Republican \\ 
   WICKER, Roger F. & Republican \\ 
   \hline
\end{tabular}
\caption{U.S. Senators Data. Names and party affiliations of the senators assigned to cluster 3 in the bottom plot of Figure \ref{fig:col_clu_sen}.}\label{table:cluster3pol}
\end{table}
\clearpage

\subsection{Movielens data}

\begin{figure}[ht!]
    \centering
    \includegraphics[height=8cm]{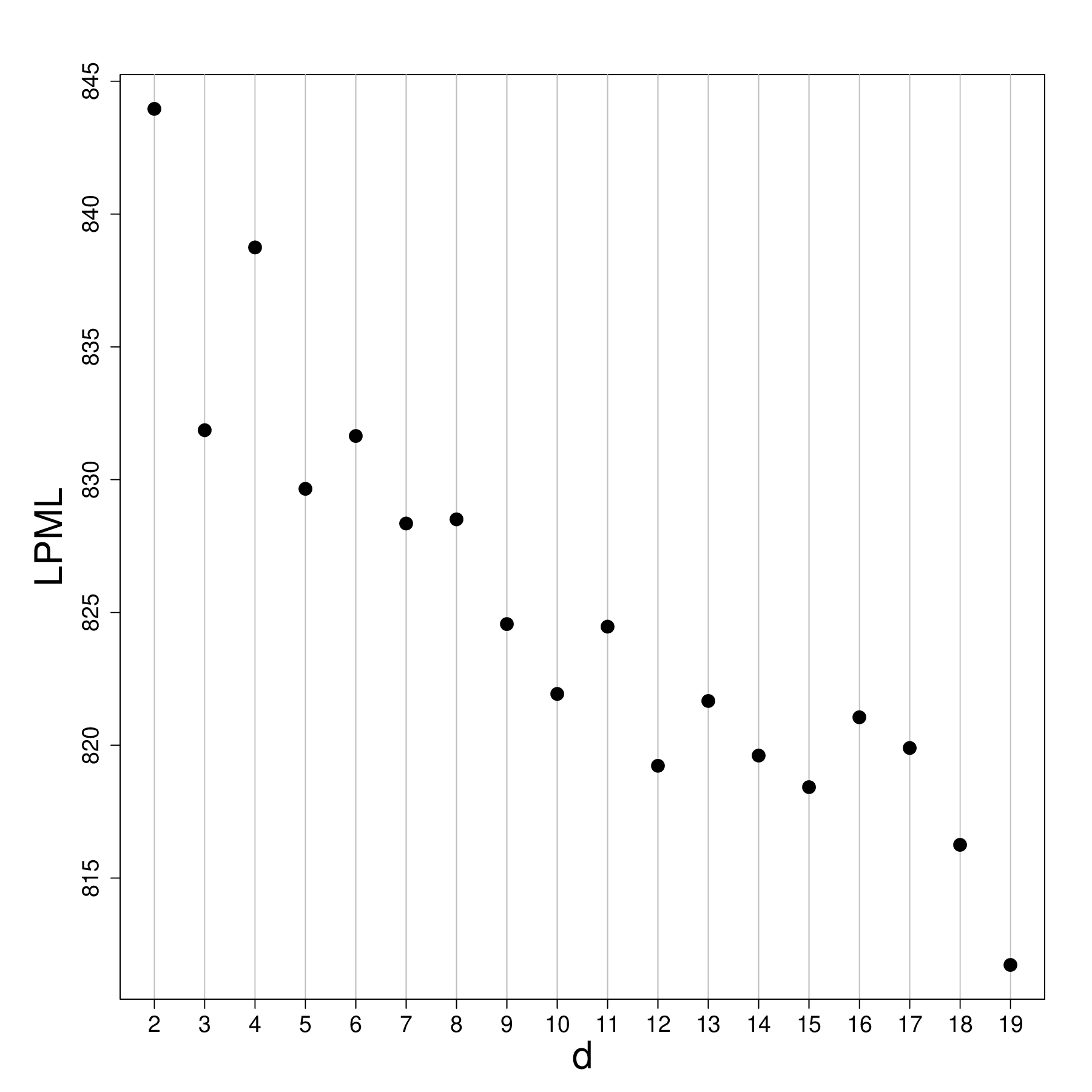}
    \caption{Movielens Data. LPML as a function of the latent dimension $d$. Refer to Section \ref{movie} for further details.}
    \label{fig:lpml_movie}
\end{figure}

\end{document}